\documentclass[twocolumn]{aastex631}
\usepackage{graphicx,amsmath, amssymb, amsthm}
\usepackage{natbib}
\usepackage{epsfig}
\usepackage{url}

\usepackage{bm}
\usepackage{color}
\begin{document}

\title{Searching for cataclysmic variable stars in unidentified X-ray sources}

\author{J..Takata}
\affiliation{Department of Astronomy, School of Physics, Huazhong University of Science and Technology, Wuhan 430074, China} 
\author{X.F. Wang}
\affiliation{Department of Astronomy, School of Physics, Huazhong University of Science and Technology, Wuhan 430074, China}
\author{A.K.H. Kong}
\affiliation{Institute of Astronomy, National Tsing Hua University, Hsinchu 30013, Taiwan}
\author{J. Mao}
\affiliation{Yunnan Observatories, Chinese Academy of Sciences, Kunming, 650216, China}
\affiliation{Key Laboratory for the Structure and Evolution of Celestial Objects, Chinese Academy of Sciences,
  Kunming, 650216, China}
\author{X. Hou}
\affiliation{Yunnan Observatories, Chinese Academy of Sciences, Kunming, 650216, China}
\affiliation{Key Laboratory for the Structure and Evolution of Celestial Objects, Chinese Academy of Sciences, Kunming, 650216, China}
\author{C.-P. Hu}
\affiliation{Department of Physics, National Changhua University of Education, Changhua 50007, Taiwan}
\author{L. C.-C. Lin}
\affiliation{Department of Physics, National Cheng Kung University, Tainan 701401, Taiwan}
\author{K.L. Li}
\affiliation{Department of Physics, National Cheng Kung University, Tainan 701401, Taiwan}
\author{C.Y. Hui}
\affiliation{Department of Astronomy and Space Science, Chungnam National University, Daejeon 305-764, Korea}

\email{takata@hust.edu.cn}

\begin{abstract}
 We carry out a photometric search for  new cataclysmic variable stars (CVs), with the goal  of identification for  candidates of   AR~Scorpii-type binary systems.  We select  GAIA sources that are likely  associated with unidentified X-ray sources, and 
analyze the light curves taken by the  Zwicky Transient Facility,  Transiting Exoplanet Survey Satellite,  and Lulin One-meter Telescope in Taiwan. We investigate  eight sources as  candidates for  CVs,  among which six  sources are new identifications.  Another two  sources have been recognized as CVs in previous studies, but no detailed investigations have been done. We identify two eclipsing systems that are associated with an unidentified XMM-Newton or Swift source, and one promising  candidate for  polar associated with an  unidentified ASKA source.
Two  polar candidates may  locate  in the so-called period gap of a CV, and the other six
candidates have an orbital period shorter than that of the period gap. Although
we do not identify a promising candidate for AR~Scorpii-type binary systems, our study suggests that CV systems that have X-ray emission and do not show frequent outbursts may have been missed in previous surveys.

\end{abstract}
\section{Introduction}
The cataclysmic variable star (hereafter CV)  is  a binary system composed of a white dwarf primary (hereafter WD)
and a low-mass main-sequence star \citep{1995cvs..book.....W}. In the usual CV system, the mass is transferred  from the companion star to the WD, and it forms an accretion disk extending down to the  WD surface or accretion column on the WD's pole, toward  which the accreting matter is channeled  by the WD's magnetic field.
The former and latter systems usually belong to   nonmagnetic and magnetic  CV systems, respectively. A nonmagnetic CV shows frequent  outbursts due to an instability of the accretion disk (dwarf nova).
Magnetic CVs, in which the WD's magnetic field is $B_{WD}>10^5$~G,
are divided into two types, namely, intermediate polar (hereafter IP)  and polar. In an IP system, the spin period of the WD 
is different from the orbital period of the system. Polar has the strongest magnetic field and shows a spin-orbit phase
 synchronization. CVs are usually observed  in the optical to X-ray bands, 
for which emission  originates from the boundary layer of the accretion disk,  the companion star surface, or  the WD surface/accretion column.

Numerous efforts to identify   new CVs and candidates have been made in previous works, and the number of known CVs is
 rapidly  increasing with recent  photometric and spectroscopic all-sky surveys \citep{1998A&AS..129...83R,  2016MNRAS.456.4441C,2021arXiv211113049S,2021AJ....162...94S}. 
The methods of confirming CVs  are mainly divided into three types, namely, the observation  of dwarf-nova outbursts, 
identification of orbital/WD spin variations in  photometric light curves,  and confirmation of CV-like spectral properties.  The Open Cataclysmic Variable Catalog \citep{2020RNAAS...4..219J}  offers a vast list  of the known CVs and  the  CV candidates found in  previous studies.

Among known WD binary systems, AR Scorpii is one of the special classes  in terms of the observed emission properties.
 The emission in the radio to X-ray bands is modulating with a spin period of WD ($\sim117$~s) and/or  a beat period
 ($\sim 118$~s)  between the WD spin and orbital motion, and its  broadband spectrum is described by a nonthermal emission process plus   thermal emission from the companion star surface \citep{2016Natur.537..374M, 2017NatAs...1E..29B,2018A&A...611A..66S,2018ApJ...853..106T}.  Although there has not yet been a direct measurement of the magnetic field of a WD \citep{2021ApJ...908..195G}, the observational properties suggest a magnetic WD binary system. The emission in the optical and X-ray bands is also modulating with the orbital period of $\sim 3.56$~hr, and the shape of the orbital light curve suggests heating of the dayside of the companion star at a rate of $L_{irr}\sim 10^{32-33}~{\rm erg~s^{-1}}$,  most of which  is converted into  emission in the IR/optical/UV bands.  The multiwavelength spectrum exhibits no features of  emission from an accretion disk in the system,  and the emission from the WD is fainter than the observed optical emission, which cannot be the source of the heating of the companion star. It is therefore suggested that  the magnetic field of the WD may  be the energy source of the heating, and it is interacting with the companion star or outflow matter from the companion \citep{2017ApJ...851..143T,2017ApJ...835..150K,2020arXiv200411474L}. AR~Scorpii  may be classified as an IP in the sense that the spin period  of the magnetic WD is shorter than the orbital period.  However, the X-ray luminosity is of the order of $L_X\sim 4\times 10^{30}~{\rm erg~s^{-1}}$, which is two to three orders of magnitude lower than that of typical IPs, in which the X-ray emission originated from the accretion column on the WD. Attention has been paid  to AR Scorpii to study  the origin of the magnetic field of 
the WD \citep{2021NatAs...5..648S,2021MNRAS.508..561W}.

An AR~Scorpii-type binary system will be  a new astrophysical laboratory for   nonthermal processes, and it may be the origin of cosmic-ray electrons. A second AR~Scorpii, however,  has not yet been identified,  and the
population in the galaxy has not been understood.  AE~Aquarii/LAMOST J024048.51+195226.9 \citep{2022MNRAS.509L..31P} are known as a  magnetic WD system in the propeller phase, and they are  similar to AR Scorpii in the sense that (i) no accretion disk is formed and (ii) the system contains a fast-spinning WD \citep{1979ApJ...234..978P,1997MNRAS.286..436W, 2021MNRAS.503.3692P,2021ApJ...917...22G}. On the other hand, there is no evidence of  heating of the companion star,  and the existence of the nonthermal emission extending in broad energy bands (radio to X-ray/gamma-ray bands) has not been established; for AE~Aquarii,  although evidence of the nonthermal emission in X-ray and TeV energy bands has been reported \citep{1994ApJ...434..292M,2008PASJ...60..387T},  the results have not been confirmed by follow-up observations \citep{2014ApJ...782....3K,2014A&A...568A.109A, 2016ApJ...832...35L}. CTCV~J2056-3014 and V1460 Her also contain  fast-spinning WD, and they are classified as X-ray faint  IPs \citep{2020ApJ...898L..40L,2020MNRAS.499..149A}. Spectroscopic studies suggest that CTCV~J2056-3014 and V1460~Her contain accretion disks and hence their WDs will have a weak magnetic field. 

AR Scorpii has  faint X-ray emission and has not shown  a dwarf-nova-type outburst. Moreover, AR Scorpii contains a 
WD with  a relatively cool surface temperature, and does not show a deep eclipsing feature in the  light curve.  Binary systems with such featureless emission properties may have  at first glance,  been missed by previous surveys, and  may require a different approach to confirm  new AR~Scorpii-type binary systems. In this study, we take the approach of finding  new CVs in unidentified X-ray sources,  since the X-ray emission as a result of the interaction between the WD's   magnetic field and the companion star is a  unique property of AR Scorpii.  The structure of this paper is as follows. Section~\ref{select} describes our strategy and method  for searching for new CV candidates. Section~\ref{result} presents 
 eight candidates including six  new identifications  and two sources that have been recognized as CV candidates but have not been  listed in the current catalogs. 
Although our new candidates will not be categorized as a new AR~Scorpii-type binary system, 
searching in the  unidentified X-ray sources offers an  alternative approach to identifying  new CV systems. In section~\ref{discuss}, we compare 
the UV and X-ray emission properties of our candidates with those of known CVs.

\section{Searching method}
\label{select}
\subsection{candidate selection}
\begin{figure}
  \epsscale{1}
  \centerline{
    \includegraphics[scale=0.5]{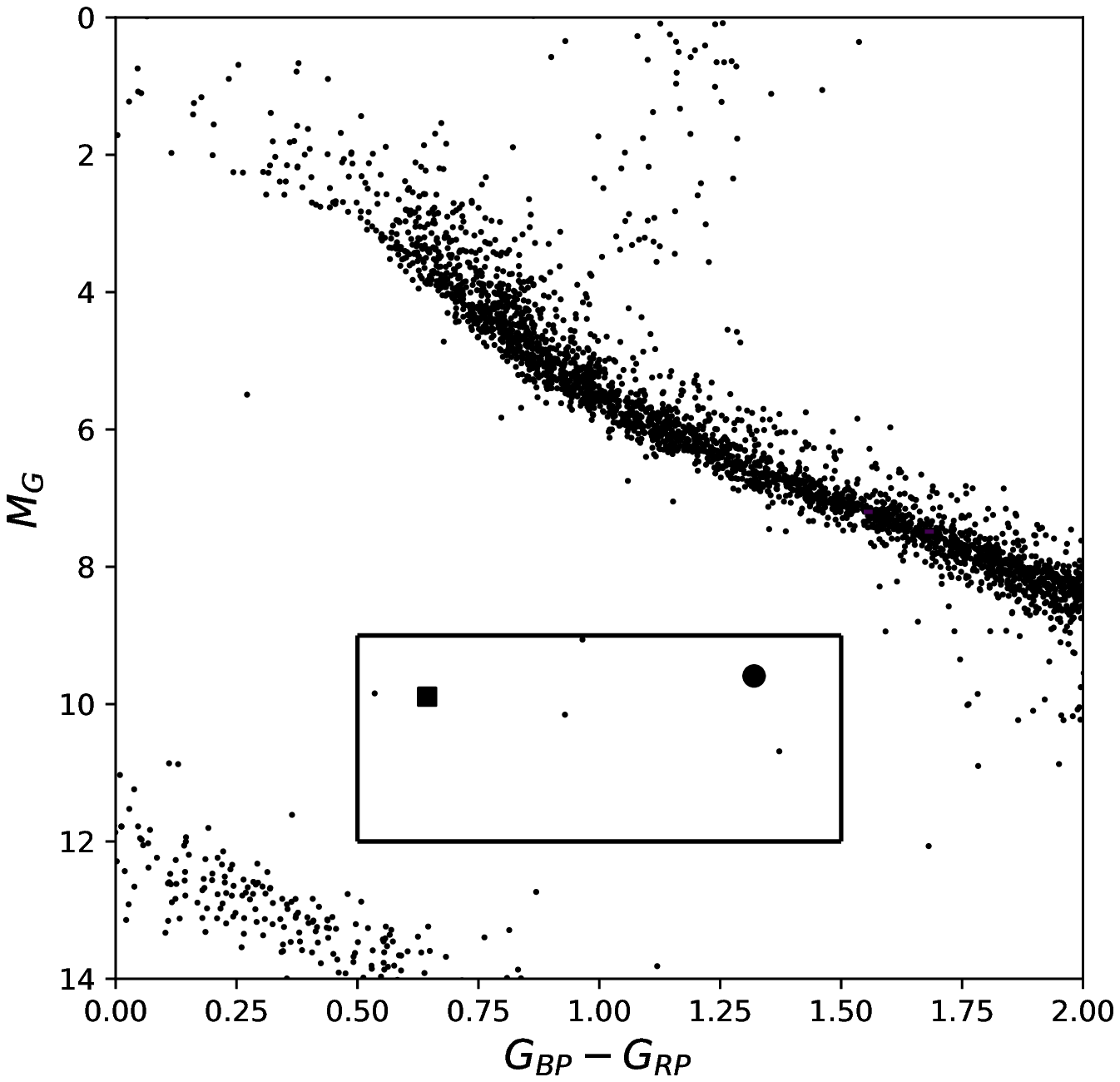}}
  \caption{Color-magnitude diagram of GAIA sources. The square represents the search region in our study. The filled circle and box correspond
    to AR~Scorpii and CTCV~J2056-3014, respectively.
  }
  \label{hr}
\end{figure}

First, we  select candidates of the WD/low-mass main-sequence star binary systems from 
the GAIA DR2 source list \citep{2018A&A...616A..10G}.
In the GAIA color-magnitude diagram,  AR~Scorpii,  CTCV~J2056-3014 and  the typical CV systems are located between the main-sequence and  the cooling  sequence of the WD (Figure~\ref{hr}).
In this work, therefore, we limit the range of the search with a color of  $0.5<G_{BP}-G_{RP}<1.5$ ($G_{BP}$ and $G_{RP}$ are blue and red magnitude defined by the  GAIA photometric system, respectively) and a magnitude of $9<M_G<12$.  We do not limit the range of the parallax, but  restrict the magnitude of the  error  with a cut \verb|parallax_over_error > 5|
in the query.   To obtain a clean sample, we refer to  \cite{2018A&A...616A...2L} and apply the conditions that 
\begin{itemize}
  \item \verb|phot_bp_mean_flux_over_error > 8|
  \item \verb|phot_rp_mean_flux_over_error > 10|
  \item \verb|astrometric_excess_noise < 1|
  \item \verb|phot_bp_rp_excess_factor < 2.0+0.06*|
    \verb|power(phot_bp_mean_mag-phot_rp_mean_mag,2)|
  \item \verb|phot_bp_rp_excess_factor > 1.0+0.015*|
    \verb|power(phot_bp_mean_mag-phot_rp_mean_mag,2)|
  \item \verb|visibility_periods_used > 5|.
\end{itemize}
We downloaded a  catalog of the GAIA sources using  \verb|astroquery| \citep{2019AJ....157...98G}.

We search for a possible  X-ray counterpart of the selected GAIA sources in  (i) the  ROSAT all-sky survey bright source catalog~\citep{1999A&A...349..389V}, (ii) the second Swift-XRT point-source catalog \citep{2020ApJS..247...54E} and (iii) XMM-Newton DR-10 source catalog~\citep{2020A&A...641A.136W}.
We select the GAIA sources that are located within 10'' from  the center of the X-ray source, and then  we remove the sources that have been
already identified as a CV or other types of objects by checking the catalogs of CVs
 \citep{2003A&A...404..301R,2016MNRAS.456.4441C,2020RNAAS...4..219J, 2021arXiv211113049S, 2021AJ....162...94S},   and the SIMBAD astronomical database\footnote{\url{http://simbad.u-strasbg.fr/simbad/}}.

After selecting the   GAIA sources that are potentially associated with unidentified X-ray sources, we  cross match them 
with sources observed by the Zwicky Transient
Facility DR-8 objects \citep[hereafter ZTF, ][]{2019PASP..131a8003M}.  We select a potential candidate of  CVs based on (i) identification of a signature of an outburst and (ii) a periodic search with a  Lomb-Scargle periodogram   \citep[hereafter LS,][]{1976Ap&SS..39..447L}.
We  download the light curves from the Infrared Science Archive  \footnote{\url{https://irsa.ipac.caltech.edu}},  and use the $r$-band data to search for a periodic signal.  We apply the  barycentric correction to the photon arrival time using \verb|astropy|\footnote{\url{https://docs.astropy.org/en/stable/time/index.html}}.    We find, on the other hand, that since the
  ZTF observation  for our candidates  (section~\ref{result}) provides only $500-1000$ data points during 2018-2021 observations,  the data quality may not be 
enough to study the  detailed properties of the light curve  (e.g. identifying eclipse feature, section~\ref{g141}). 

The Transiting Exoplanet Survey Satellite~\citep[hereafter TESS,][]{2014SPIE.9143E..20R} provide the light curves of the sources by monitoring for a month, and also photometric data of  sources  that are out of
the field of view of ZTF. For our targets, TESS full-frame images (hereafter FFIs) provides the data taken every  10~minutes or  30 minutes, for which Nyquist frequencies are $F_{N}\sim 72~{\rm day^{-1}}$ and $\sim 24~{\rm day^{-1}}$, respectively.  We extract the light curve of the pixels around  the source region from TESS-FFIs  using TESS analysis tools, \verb|eleanor| \citep{2019PASP..131i4502F,2019ascl.soft05007B} and \verb|Lightkurve| \citep{2018ascl.soft12013L}.  We note that since several  GAIA sources 
 are usually located at one pixel, TESS data alone cannot tell us which source produces 
the  periodic signal in a light curve extracted from TESS-FFIs. To complement ZTF and TESS observations, we carry out photometric observations for some of our targets with the  Lulin One-meter Telescope (hereafter LOT) in Taiwan.

\subsection{LS periodogram}

First, we produce an LS periodogram for each source by taking into account the  Gaussian and uncorrelated  errors that are usually provided
in the archival data (or by usual data processing). We search for a possible periodic signal in $1~{\rm day^{-1}}<f<50-150~{\rm day^{-1}}$, in which the maximum frequency depends on the time resolution of the observation. We estimate the false alarm probability (hereafter FAP) of the signal with the methods  of  \cite{2008MNRAS.385.1279B} and of the bootstrap~\citep{2018ApJS..236...16V}. For an accreting system, it is proposed to  investigate the  effect of the time-correlated noise on the LS periodogram~\citep{2018ApJS..236...16V}. Based on the correlated noise model of  \cite{2020A&A...635A..83D}, we, therefore, produce  an LS periodogram with
different parameters of  the noise model (Appendix~\ref{noise}). We find that
the LS periodogram of TESS data is insensitive to time-correlated noise model. For ZTF data, although the noise model can change the shape of the LS periodogram, it is  less effective in  the periodic signals presented in this study (Table~1); but see section~\ref{g141} and Appendix~\ref{noise} for ZTF18aampffv,
in which one  periodic signal may be  related to time-correlated noise.
In section~\ref{result}, therefore, we present an LS periodogram
created with the time uncorrelated noise. The correlated noise model and
some results of the LS periodogram are presented in Appendix~\ref{noise}.

\section{Results}
\label{result}

\begin{deluxetable*}{ccccccc}
\tablecolumns{7}
\tabletypesize{\footnotesize}
\tablecaption{Basic information on the CV candidates in this study}
\tablehead{
\colhead{GAIA}  & 
\colhead{ZTF}&
\colhead{X-ray source} &
\colhead{Distance}  &
\colhead{$f_{ZTF}/F_0$\tablenotemark{\rm a}} &
\colhead{Orbital} &
\colhead{Proposed type} \\
\colhead{DR2} & 
\colhead{}& 
\colhead{}&
\colhead{(pc)}  &
\colhead{(${\rm day^{-1}}$)} &
\colhead{frequency}  &
\colhead{}
 }
\startdata
    1415247906500831744 (G141) & 18aampffv &4XMM J172959.0+52294
    & 532  & - /11.148(4)  & $F_0$  & Eclipsing\\
    4534129393091539712 (G453) & 18abikbmj &1RXS J185013.9+242222   & 322  & - / 39.11(4) & $F_0$/2 or $F_0$ & Dwarf-Nova/superhump    \\
    2072080137010352768 (G207) & 18abrxtii &2SXPS J195230.9+372016& 313 & 28.575(1)/28.57(4) & $F_0$/2  & Eclipsing \\
    2056003803844470528 (G205) &18aayefwp  &2SXPS J202600.8+333940&  595  & 24.595(1)/24.60(4) & $F_0$/2 or $F_0$ &\\
    2162478993740496256 (G216) & 17aaapwae & 2SXPS J211129.4+445923& 429  & 22.175(1)/22.18(2) & $F_0$ &  \\
    4321588332240659584 (G432)& 18aazmehw & 2SXPS J192530.4+155424 & 582  & 18.5521(9)/18.553(1)& $F_0$/2 or $F_0$ & IP\\
    4542123181914763648 (G454)&  18abttrrr & 1RXS J172728.8+132601 & 502  & 8.6499(8)/8.65(2) & $F_0$ & Polar 
\enddata
\tablenotetext{\rm a}{$f_{ZFT}$ and $F_0$ correspond to photometric periodic signals seen 
in the ZTF and TESS light curves, respectively.}
\end{deluxetable*}

\begin{deluxetable*}{cccccc}
  \tablecolumns{6}
  \tabletypesize{\footnotesize}
  \tablecaption{Information on  TESS and  LOT observations}
  \tablehead{
    \colhead{GAIA}  &
    \colhead{TESS}&
    \colhead{} &
    \colhead{LOT} &
    \colhead{} &
    \colhead{Figures\tablenotemark{\rm a}}
    \\
    \colhead{DR2} &
    \colhead{Date (MJD)}&
    \colhead{Sector}&
    \colhead{Date (MJD)}  &
    \colhead{Exposure (hrs)} & 
    \colhead{}
  }
\startdata
  G141 & 58764-58789 &17   & 59491/59492  & 5.2 & 2-5, A1 \\
  & 58954-59034 & 24, 25, 26&        & &\\
  & 59579-59606 & 47 & && \\
  & 59664-59690 & 50 & && \\
  \hline
  G453 & 59009-59034 & 26   & 59398/59399 & 4.9 & 6-9 \\
\hline
  G207 & 58683-58736 & 14, 15& 59475/59476  &2.8 &10,11\\
  & 59419-59445 & 41  & & &\\
\hline
  G205 & 58683-58736& 14, 15&  59542 & 0.7& 10, A1 \\
  & 59419-59445 & 41  & & &\\
  \hline
  G216 & 58710-58762& 15, 16 &59474 & 2.7 & 10,12  \\
\hline
  G432& 58682-58709 & 14 & & & 13,14\\
 & 59390-59418&  40 & & &\\
\hline
  G454& 58984-59034 &  25, 26 & &  & 15 
  \enddata
 \tablenotetext{\rm a}{References for figures in this paper.}
  
\end{deluxetable*}

\begin{deluxetable*}{cccccc}
\tablecolumns{6}
\tabletypesize{\footnotesize}
\tablecaption{Summary of Swift X-ray observations}
\tablehead{
  \colhead{GAIA}&
   \colhead{Data} &
  \colhead{Date/Exposure}&
  \colhead{$N_H$}  &
  \colhead{Photon index} &
  \colhead{Luminosity}
  \\
  \colhead{DR2} &
  \colhead{}&
    \colhead{(MJD)/(ks)}&
  \colhead{$10^{22}(\rm{cm^2})$}  &
  \colhead{} &
  \colhead{$10^{31}({\rm erg~s^{-1}})$}  
}
\startdata
G141 & TOO & 59558/4.7 & $1.9^{+5.4}_{-1.9}$&  $0.5^{+0.6}_{-0.5}$ & $4.5^{+1.9}_{-1.3}$  \\
G453 & TOO & 59529, 59534/2.0 &$1.4^{+2.5}_{-1.4}$  & $1.6^{+0.7}_{-0.6}$&   $5.5^{+2.0}_{-1.5}$ \\
G207 & Archive & 57046, 57047/3.2 & 0.3 (fixed)\tablenotemark{\rm a}  & $0.9^{+1.0}_{-1.0}$& $0.5^{+0.8}_{-0.3}$ \\  
G205 & Archive & 57796-57806/9.1 &$4.0^{+5.2}_{-2.0}$ & $2.4^{+1.1}_{-0.9}$ &  $1.4^{+3.4}_{-0.6}$ \\
G216 & Archive  & 57210-57309/10 & 4.0 (fixed)\tablenotemark{\rm a} & $1.8^{+0.4}_{-0.4}$& $1.2^{+0.4}_{-0.3}$ \\
G432& Archive  &56011-56074/1.1 & 13 (fixed)\tablenotemark{\rm a} & $0.3^{+1.1}_{-1.4}$ & $14^{+24}_{-8.0}$\\
G454&  TOO  & 59591/2.5  &  0.75 (fixed)\tablenotemark{\rm a}& $-0.2^{+1.0}_{-1.6}$ & $5.5^{+14}_{-3.2}$
\enddata
\tablenotetext{\rm a}{$N_H$ is estimated from the sky position using  the hydrogen column density calculation tool ``${\rm N_H}$''  under HEASoft, and is fixed during the fitting process.}
\end{deluxetable*}

We downloaded   the catalog of $\sim 2\times 10^5$ GAIA sources selected based on criteria described  in section~\ref{select}, and
identify 29 sources that are potential counterparts of the unidentified X-ray sources.  After searching for period
signals in the light curves  based on ZTF, TESS, and LOT data, 
we identify  seven  sources as being CV candidates (Tables~1-3, sections~\ref{g141}-\ref{g452}), for which we can obtain a periodic signal with  at least two different facilities. Two  of them  have been recognized as candidates for CVs in previous studies, but they are not listed in the current catalogs.  We also present four  other sources
(Table~4, section~\ref{others}), for which no useful ZTF data are available, but TESS-FFI data indicates a periodic signal, and one  of them is a promising candidate for polar.  LS periodograms and light curves obtained from
ZTF and TESS data for 11 candidates are presented in Figures~\ref{ztf}-\ref{other}.

\subsection{GAIA DR2 1415247906500831744}
\label{g141}


\begin{figure*}[htb]
  \centering
  \begin{tabular}{cc}
    {\includegraphics[width=0.3\linewidth]{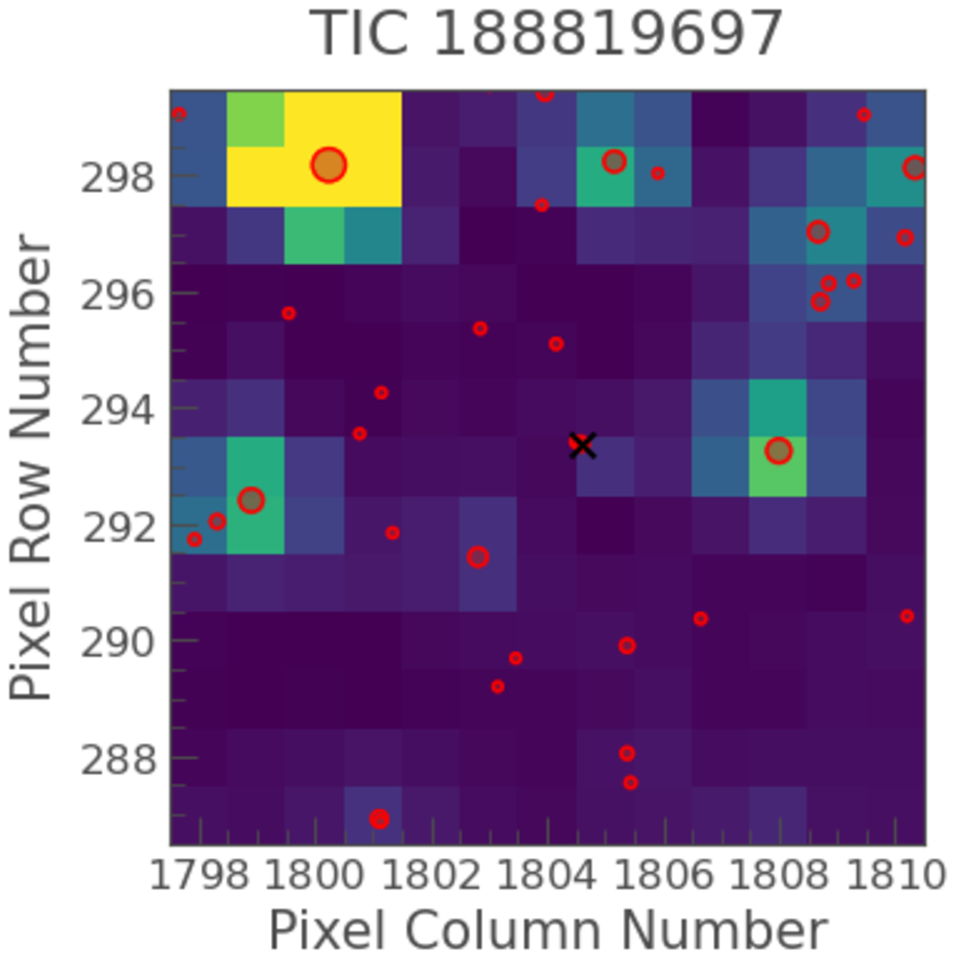}} &
    {\includegraphics[width=0.4\linewidth]{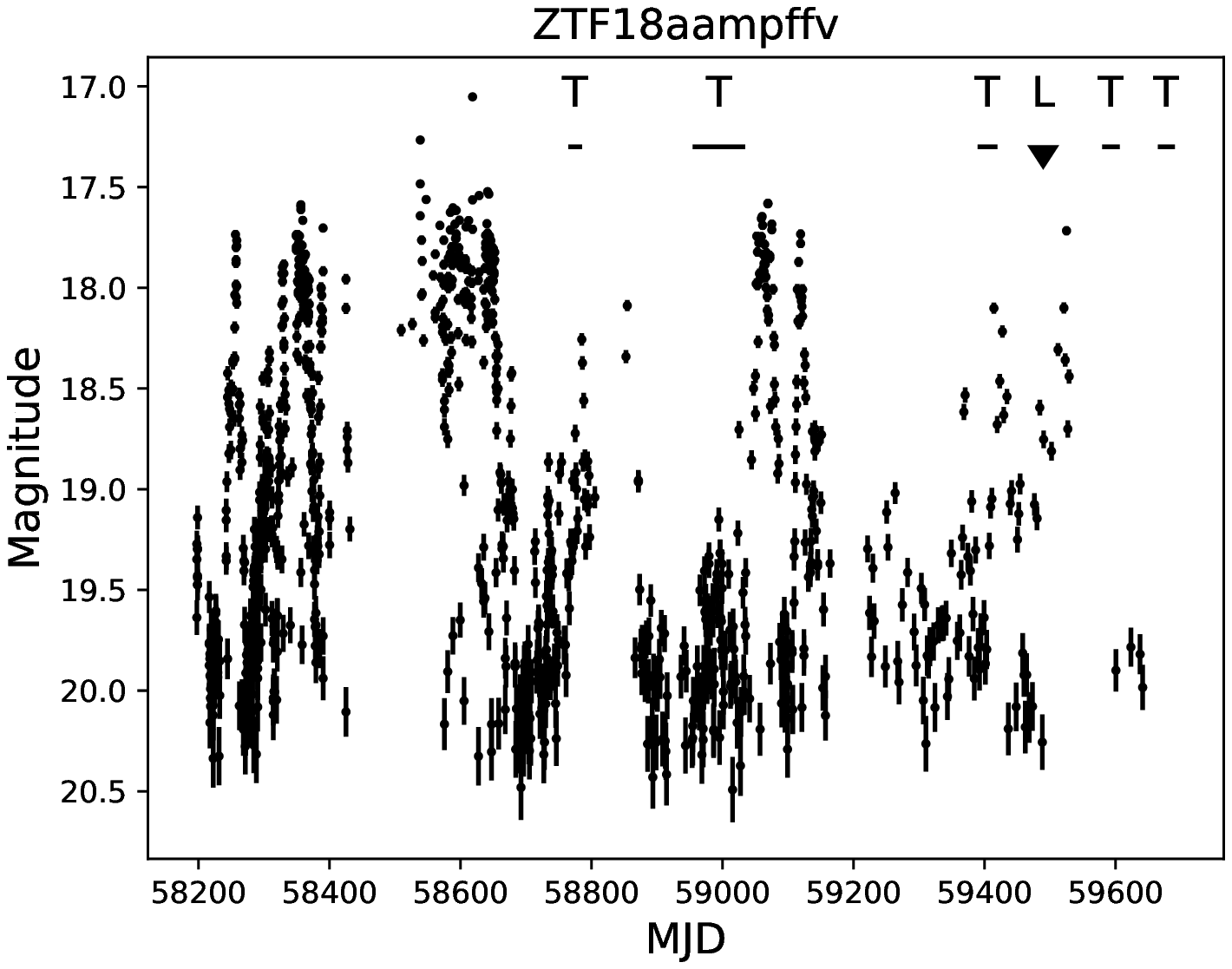}} \\
    {\includegraphics[width=0.4\linewidth]{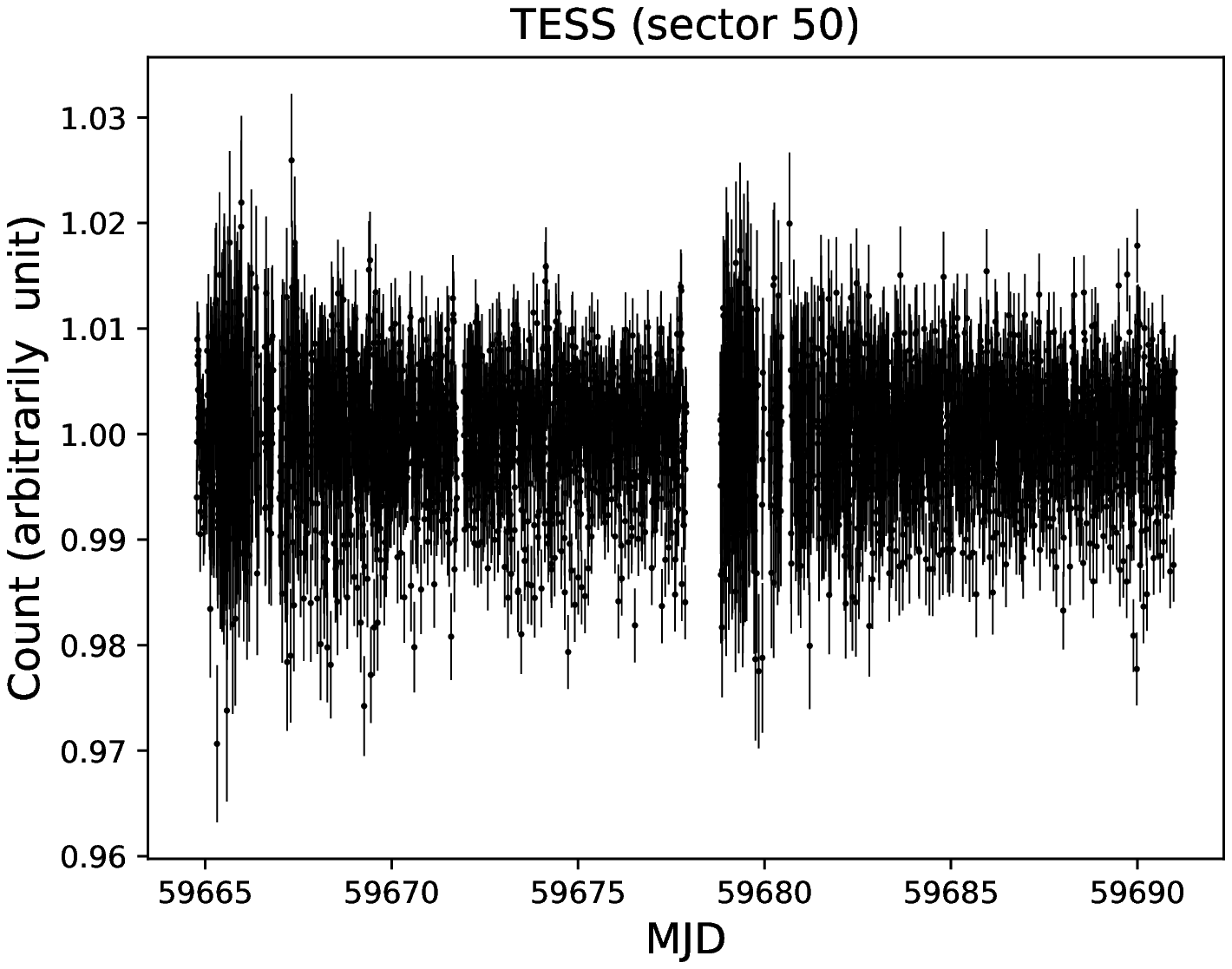}} &
    {\includegraphics[width=0.4\linewidth]{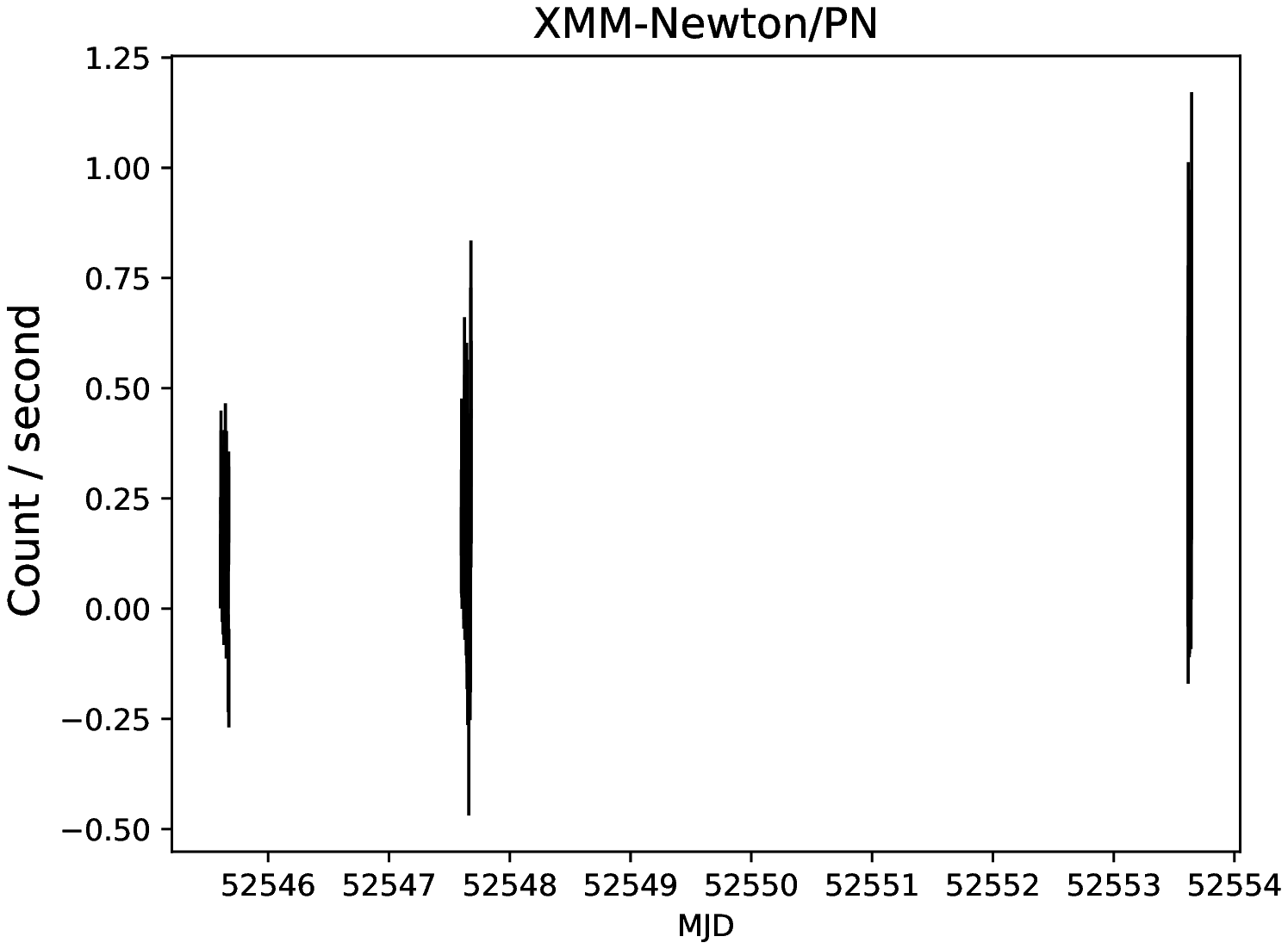}}
  \end{tabular}
  \caption{ Top left: TESS-FFI image of G141. The cross symbol and red circles indicate the position of
    the target and the GAIA sources with  a magnitude $m>20$, respectively.
    Top right : ZTF light curve  (g-band) of the target (ZTF18aampffv).
    ''T''(horizontal bars) and ``L'' (triangle) indicate the epochs of TESS  and LOT observations,
    respectively. Bottom left :  Light curves taken by TESS (sector 50). The light curve extracted
    from TESS-FFI is corrected with TESS analysis tool {\tt Lightcurve}\citep{2018ascl.soft12013L}.
    Bottom right: Light curve with  XMM-Newton PN data. }
    \label{ztf18aam}
  \end{figure*}

\begin{figure*}
  \epsscale{1}
  \centerline{
    \includegraphics{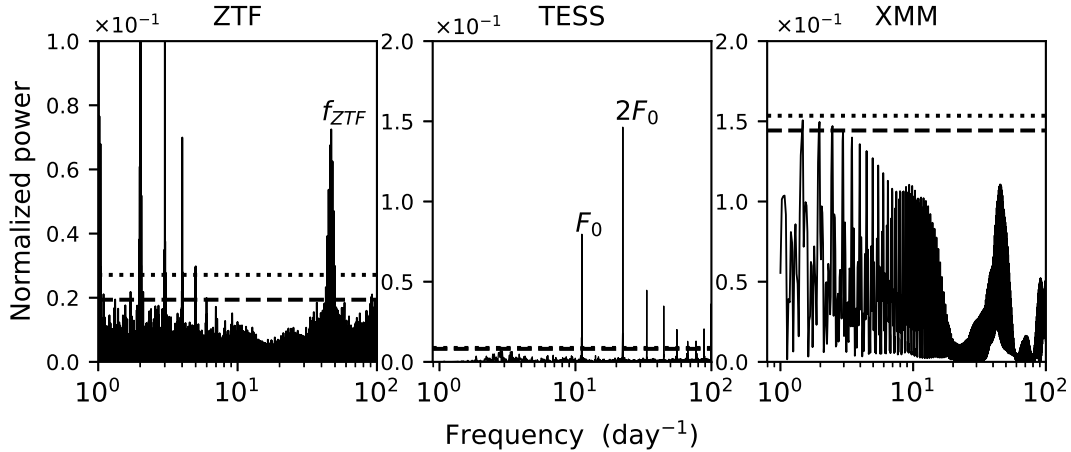}} 
  \caption{LS periodiagrams of ZTF (left), TESS (middle, sector~50) and XMM-Newton PN (right) data for G141.  The dotted and dashed horizontal lines represent the FAP=0.01 estimated
    by the methods of \cite{2008MNRAS.385.1279B} and of the bootstrap~\citep{2008MNRAS.385.1279B}. In the TESS data (middle panel), $F_0$ represents the orbital period of this system. } 
  \label{power}
    \end{figure*}

\begin{figure*}
  \epsscale{1}
  \centerline{
    \includegraphics[scale=1.1]{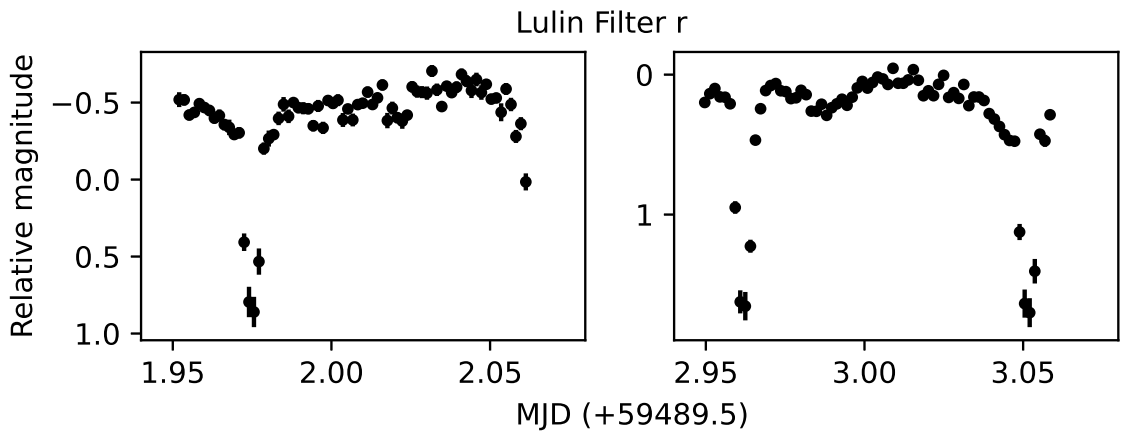}
  }
  \caption{Light curve of G141 taken by LOT. The observation covers three eclipses
    and confirms an orbital frequency of $F_0\sim 11.148~{\rm day^{{-1}}}$. In the left panel, the secondary eclipse  with a  smaller dip can be seen at MJD 2.025+59489.5.}
  \label{18aam_BJD0405}
\end{figure*}

\begin{figure}
  \epsscale{1}
  \centerline{
    \includegraphics[scale=0.7]{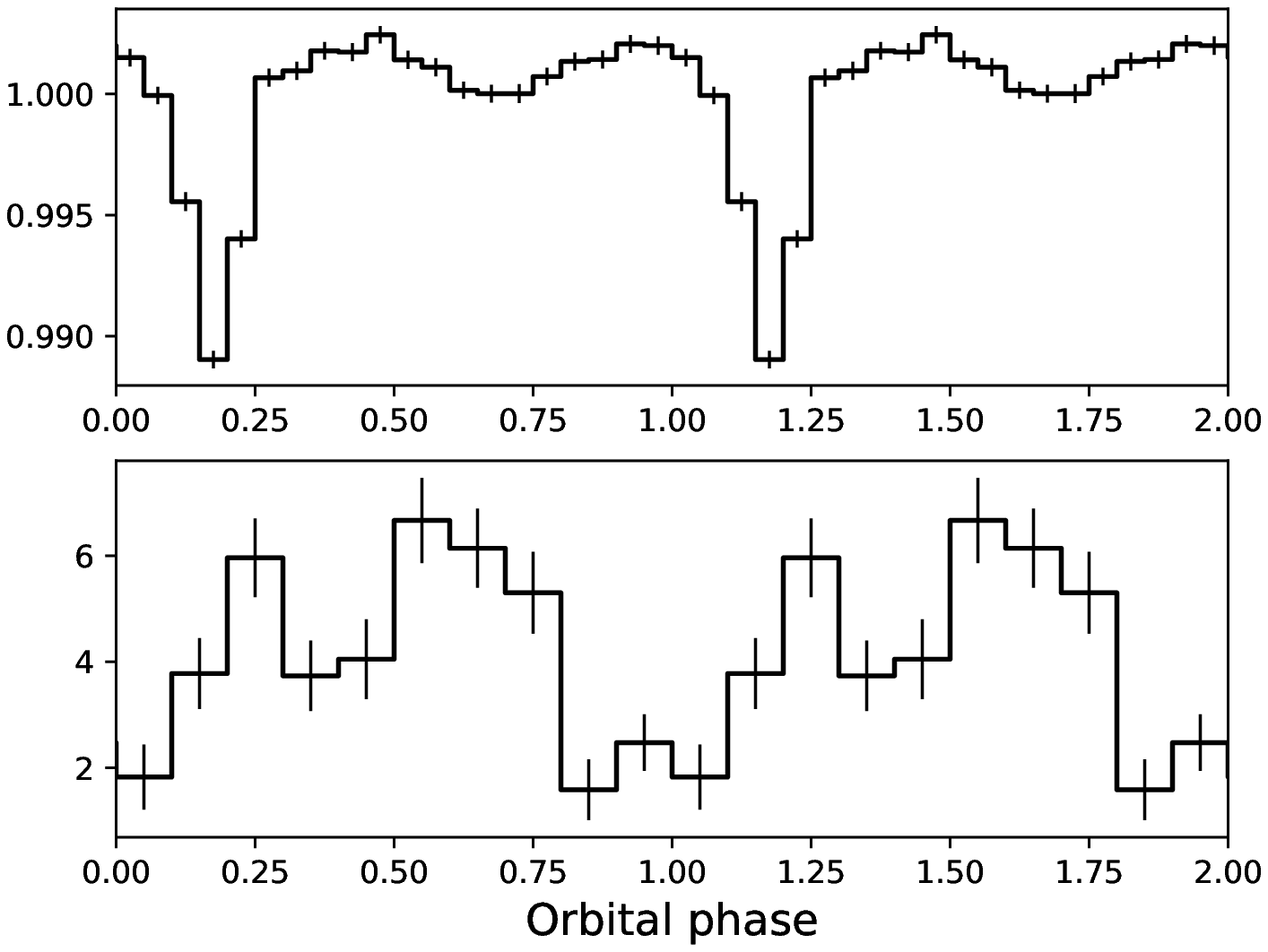}
  }
  \caption{Orbital light curve of G141 measured by TESS (top) and XMM-Newton/PN (bottom). In the TESS light curve,  two eclipses can be  clearly seen. The data is folded with $F_0=11.148~{\rm day^{-1}}$ ($\sim 0.090$~day), and  phase zero corresponds to MJD~59398.713391.}
  \label{18aam-orbit}
\end{figure}

4XMM J172959.0+522948 (hereafter, 4XMMJ1729) is an unidentified X-ray source, and its  position  is consistent with  GAIA DR2 1415247906500831744 (hereafter G141, Figure~\ref{ztf18aam}).  Based on the parallax measured by GAIA, the distance to this source  is $\sim 532$~pc. There is a nearby source (GAIA DR2 1415247906499736448), whose position from our target is  separated by $\sim2''$. Since the GAIA G-band mean magnitude of the nearby source, $\sim20$, is larger than  G141 ($\sim 18.3$ mag), contamination will have less influence on the optical light curve observed  for our  target.

Figure~\ref{ztf18aam} (top right panel) shows  the $r$-band light curve for G141  measured by ZTF (ZTF18aampffv). We find in the figure that the target  frequently repeats small outbursts or  transitions between high and low states on a timesale of months, which may indicate an accreting system. The amplitude is $\delta m > 2$, which is relatively large compared to those  seen in the ZTF light curves for other candidates presented in this study. We search for a potential periodic signal in the ZTF light curve  with the  LS periodogram (left panel of Figure~\ref{power}),
and find a significant signal at $f_{ZTF}\sim  46.9991(9)~{\rm day^{-1}}$ ($\sim 0.021$~days), where the  uncertainty  denotes the Fourier resolution of the observation (namely, the inverse of the time span covered by the  observation).

The TESS observation covered  the source region several times (Table~2).
  The middle panel of Figure~\ref{power} shows an LS periodogram of the light curve of  the TESS data, the  sector 50 (Figure~\ref{ztf18aam}). The time resolution of the sector~50 is about 10 minutes, and it has the capability  to searching for a signal of $f_{ZTF}\sim 47~{\rm day^{-1}}$. We find,
  however, that the LS periodogram is dominated by the signal with $\sim 22~{\rm day^{-1}}$, which is indicated with $2F_0$ in the figure, its harmonics and aliasing signals. We  check the data of sector 47, which  also does not show a significant periodic signal at  $f_{ZTF}$.
  Since the data of sectors 17 and 24-26 were taken approximately every $30$~minutes, they cannot be used to  search for  a periodic signal of $f_{ZTF}$. The signal of $\sim 22~{\rm day^{-1}}$ is confirmed in all data (although  sector~40 also covers the source,  the data is contaminated by  unusually high background emission or noise).

  To collect  evidence of a binary nature, we carried out a photometric observation with  LOT (2021, October 4th and 5th, Table~2).
  The light curve  clearly shows that G141 is an eclipsing binary system  (Figure~\ref{18aam_BJD0405}),
and the observation  covered  three eclipse events that repeat at  a frequency of $F_0=11\pm 4~{\rm day^{-1}}$, suggesting that G141 is  a binary system with an orbital period of $P_{orb}\sim 0.09$~day. With the time  resolution of $\sim$2 minutes for the LOT data,  we estimate that each  eclipse lasts  $\sim15$~minutes, and the eclipse profile does not change during the observation.  With the LOT data, we unable to confirm a  periodic  signal corresponding to $f_{ZTF}\sim 47~{\rm day^{-1}}$.

The top panel of Figure~\ref{18aam-orbit} shows the folded light curve of the TESS data with a frequency of
$F_0\sim 11~{\rm day^{-1}}$, and clearly  shows an eclipsing feature at the orbital phase of $\sim 0.2$ in the light curve.  We also find the  secondary eclipse around the orbital phase $\sim 0.7$, which
is shallower than that of the primary eclipse. This illustrates that the signal of the second harmonic ($F_1=2F_0\sim 22.296~{\rm day^{-1}}$) is stronger than that of the fundamental signal in the LS periodogram.


4XMMJ1729 was covered  by the field of view, when  XMM-Newton observations  were carried out  for two stars (HD 150798 and HD 159181) in 2002.  We extract the data in the  standard way using the most updated instrumental calibration, tasks \verb|emproc| for MOS and \verb|epproc| for PN of  the  XMM-Newton Science Analysis Software (XMMSAS, version 19.0.1). Although the observation  was carried out  with a total exposure  of about 20~ks, the quality of the data is insufficient to measure
the spectral properties, since (i) the target is located at the edge of the field of view in all EPIC data and (ii) the observation is significantly  affected
by background flare-like events.  We create
an LS periodogram for the  light curve (Figure~\ref{ztf18aam}) of PN data;
we do not analyze   MOS data due to an insufficient count rate. As can be seen in Figure~\ref{power}, the window effect
dominates the periodogram, and  a significant periodic signal at  $f_{ZTF}\sim  47~{\rm day^{-1}}$ is not identified, although an indication may exist. Figure~\ref{18aam-orbit} shows the 
X-ray light curve folded with the orbital period $F_0=11.148~{\rm day^{-1}}$. Although
a lower count rate is observed at $\sim0\sim 0.3$ orbital phase,
a deeper  observation is required to investigate the eclipse feature in X-ray bands.

We obtain an  $\sim 5$~ks Swift observation for 4XMMJ1729 (Table~3). We extract a cleaned event file with  \verb|Xselect| under \verb|HEASoft ver.6-29|, and fit the spectrum with  \verb|Xspec ver.12.12|. Since we could not constrain the spectral model because of insufficient photon counts,  we fit the spectrum with
a single power-law function (Table~3). Using the distance measured by GAIA (Table~1), we estimate the luminosity in the  0.3-10~keV band as
$L_X=4.5^{+1.9}_{-1.3}\times 10^{31}(d/{\rm 532~pc})^{2}~\rm{erg~s^{-1}}$ (hydrogen column density of $N_H=1.9^{+5.4}_{-1.9}\times 10^{21}~{\rm cm^{-2}}$ and photon index of $\Gamma=0.5^{+0.6}_{-0.5}$).

 We have not found a  significant periodic signal with  $f_{ZTF}\sim  47~{\rm day^{-1}}$ in TESS, LOT and XMM-Newton data.
  In Appendix~A, therefore, we carry out a further investigation of the LS  periodogram of the ZTF data and find the  possibility that the signal is caused by the time-correlated noise.

\subsection{GAIA DR2 4534129393091539712}

\begin{figure*}
  \epsscale{1}
  \centerline{
    \includegraphics[scale=0.55]{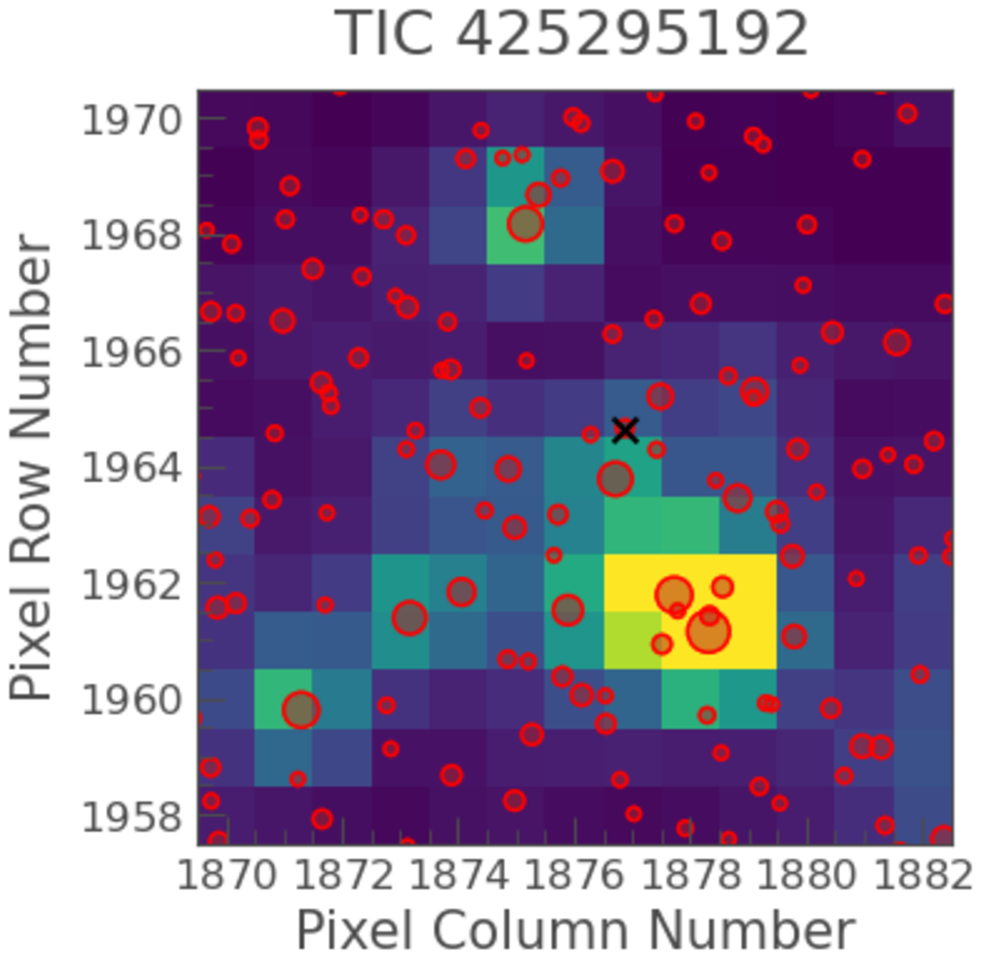}
    \includegraphics[scale=0.5]{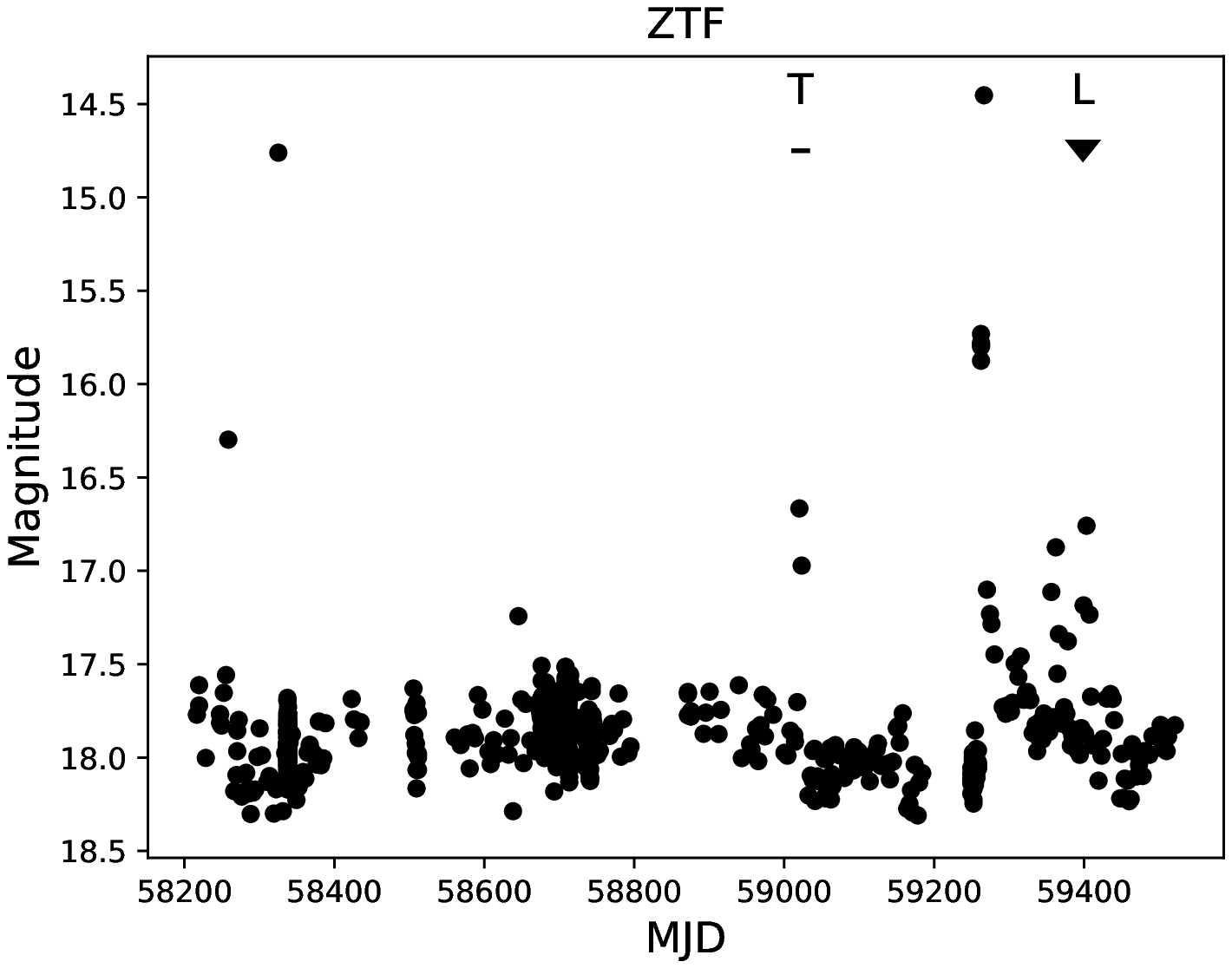}
  }
  \caption{Position (left) and light curve (right) of G453. 
Left:  TESS FFI in the source  region  overlaid by the GAIA sources (red circles) with a magnitude
    $m>20$. The center (cross symbol) represents the position of the target. Right: ZTF light curve of the 
target (ZTF18abikbmj).}
  \label{ztf18abi}
\end{figure*}

\begin{figure*}
  \epsscale{1}
  \centerline{
    \includegraphics{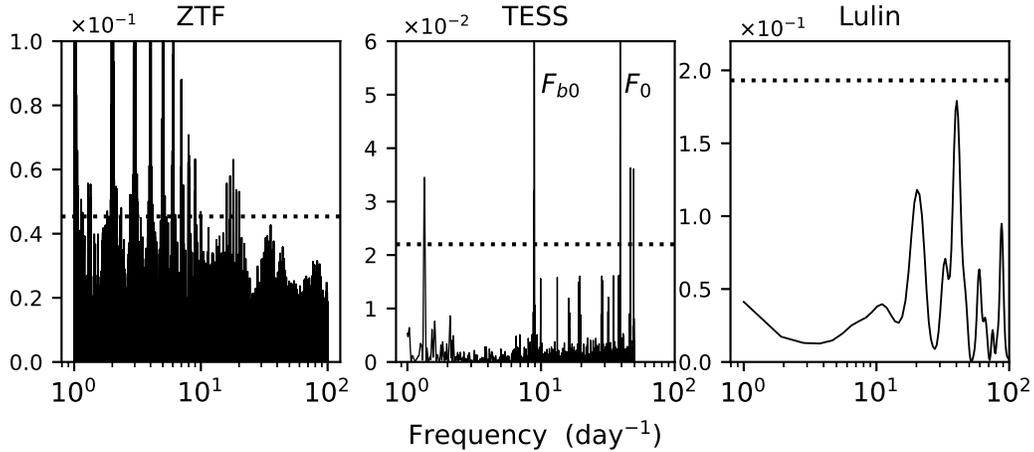}} 
  \caption{LS periodiagrams of ZTF (left), TESS (middle) and LOT (right) data for G453.  The dotted lines show the  FAP=0.01 estimated
  by the bootstrap method. }
  \label{power-abi}
    \end{figure*}
\begin{figure}
\epsscale{1}
\centerline{
  \includegraphics[scale=0.7]{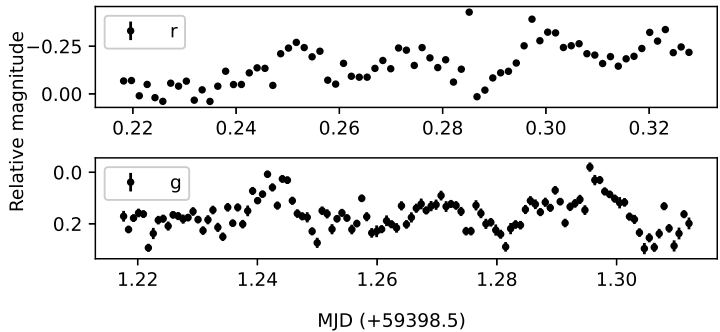}
 }
 \caption{The light curves with the $r$-band (upper) and $g$-band (lower) filters of  G453 taken by LOT. }
\label{18abi-BJD0304}
\end{figure}

\begin{figure}
  \epsscale{1}
  \centerline{
    \includegraphics[scale=1]{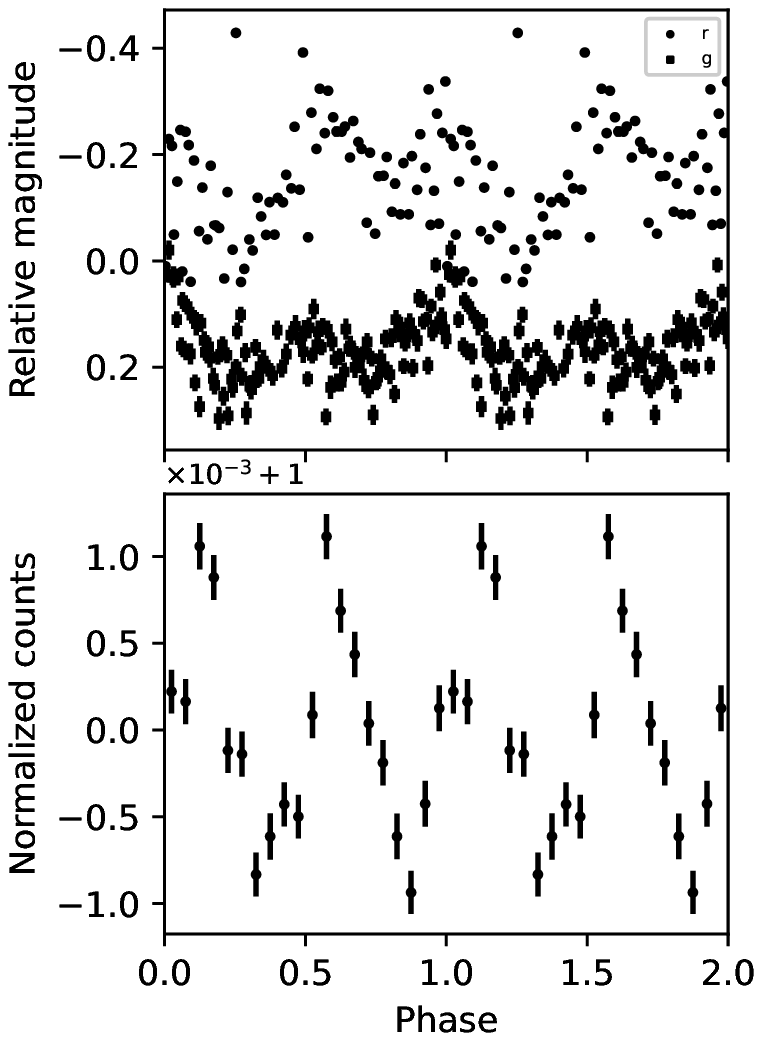}
  }
  \caption{Light curve of G453 folded  with $F_0/2=19.6~{\rm day^{-1}} (\sim 0.05~{\rm days})$. Top: Light curves with  the $r$-band (circle)  and
    $g$-band (square) taken by LOT. Bottom: Light curve taken by TESS. For TESS data, the counts are binned and each bin covers a  0.1 phase.  Phase zero refers to MJD~59398.713391.
}
  \label{abi-fold}
  \end{figure}

GAIA DR2 4534129393091539712 (hereafter, G453) is selected as a possible  counterpart of the ROSAT source, 1RXS~J185013.9+242222. We obtain an $\sim2$~ks Swift observation for 1RXS J185013.9+242222 in 2021 November,   and we estimate the  luminosity of $L_X\sim 5.5^{+2.0}_{-1.5}\times 10^{31}(d/322~\rm{pc})^2~{\rm erg~s^{-1}}$  in the 0.3-10~keV band by fitting the spectrum with a power-law model ($N_H=1.4^{+2.5}_{-1.4}\times 10^{21}~{\rm cm^{-2}}$ and $\Gamma=1.6^{+0.7}_{-0.6}$, Table~3).

As can be seen in the right panel of  Figure~\ref{ztf18abi}, the ZTF light curve (ZTF18abikbmj) shows several outbursts  with a recurrent time scale of a  year.  In fact,  1RXS~J185013.9+242222 has already been recognized as a dwarf nova  by previous observation of the outburst \footnote{\url{http://ooruri.kusastro.kyoto-u.ac.jp/mailarchive/vsnet-alert/25434}}.  Since no detailed investigation for this source  has been carried out, we searched for possible orbital modulation  in the photometric data.

First, we searched for periodic signals in a quiescent state in the ZTF data, but we did not find  any obvious periodic signals (except for the window effects)  in the periodogram (Figure~\ref{power-abi}). TESS observed this source in 2020 June during a stage of the small outburst (Figure~\ref{ztf18abi}), and  
the data was taken approximately every  $30$~minutes. We extract the light curve with 4 pixels around the target (left panel of Figure~\ref{ztf18abi}).
In the LS-periodogram, we find two strong signals at 8.88(4) day$^{-1}$ ($\sim 0.11$~days)
and 39.11(4) day$^{-1}$ ($\sim 0.026$~days), either of which is likely the aliasing of the sampling frequency. Since the extracted light curves with 4 pixels contain several sources (left panel of Figure 7), the TESS observation alone cannot support  the detected signal originating from G453.

We confirm the binary nature  of  G453 with LOT. We carried out the observations with $r$-band (2021 July 3) and g-band (2021 July 4)  filters, for which  one observation covered the source with an exposure of  $\sim 2-2.5$ hr (Table~2). As 
indicated in Figure~\ref{ztf18abi}, the LOT observation was also carried out during a  small outburst, which could be a re-brightening after the large outburst happen around MJD~59300.   In the  LS-periodogram (Figure~\ref{power-abi}) and the
  observed light curve (Figure~\ref{18abi-BJD0304}),  we  confirm evidence of  periodic modulation at
  $F_0=37.3\pm 1~{\rm day^{-1}}$, which is close to the TESS result.

Although we cannot firmly conclude the origin of the periodic signal with the current photometric data, the signal $F_0$ is likely related to the orbital period. From the shape of the light curve taken by LOT, we can expect that the orbital period is double the photometric period, namely  $P_{orb}\sim 2\times 1/F_0\sim 0.05~{\rm days}$; e.g. the light curve with  the $r$-band filter may be more consistent with modulation caused by the elliptical shape of  the companion star that fills the Roche-lobe, and an indication of
double-peak structure with the different peak magnitudes in the $g$-bands (Figure~\ref{abi-fold}).

Another possible origin of the signal $F_0$ is the superhump, which is a periodic variation observed emission from an eccentric disk   after the outburst of CVs \citep{1995cvs..book.....W}. The period of the superhump is several percent longer (or sometimes shorter) than the orbital period.  Although  superhump is usually observed after a super-outburst of CVs \citep{2005PJAB...81..291O, 2014PASJ...66...90K,2020AdSpR..66.1004H}, it has also been observed   at normal outburst or re-brightening  after a  super-outburst  \citep{2006PASJ...58..367Z,2012PASJ...64L...5I}. The ZTF light curve of G453 suggests that there were several re-brightenings after the large outburst occurred  around MJD~59300, and the LOT observation covered one of re-brightening stages. Moreover, the previous observations for other dwarf novae confirm a beat  signal between the orbital modulation and superhump \citep{2000PASP..112.1567P,2010PASJ...62.1525K}. In the periodogram of the TESS data for G453 (middle panel of Figure~\ref{power-abi}), we notice a periodic signal at $\sim 1.34~{\rm day^{-1}}$ ($\sim 0.75$~day), which is confirmed only at the source region,  with a  significance greater than the 99~\% confidence level.  This signal can be explained by the beat signal, if the difference between the orbital period and the superhump period is $\epsilon \sim 3-4$~\% (i.e. $1.34~{\rm day^{-1}}\sim \epsilon \times 39~{\rm day^{-1}}$). The TESS light curve folded with $F_0$ is presented in Figure~\ref{all-light}.


\subsection{GAIA DR2~2072080137010352768, 2056003803844470528, and 2162478993740496256}
\label{threes}
\begin{figure}
  \epsscale{1}
  \centerline{
    \includegraphics[scale=0.55]{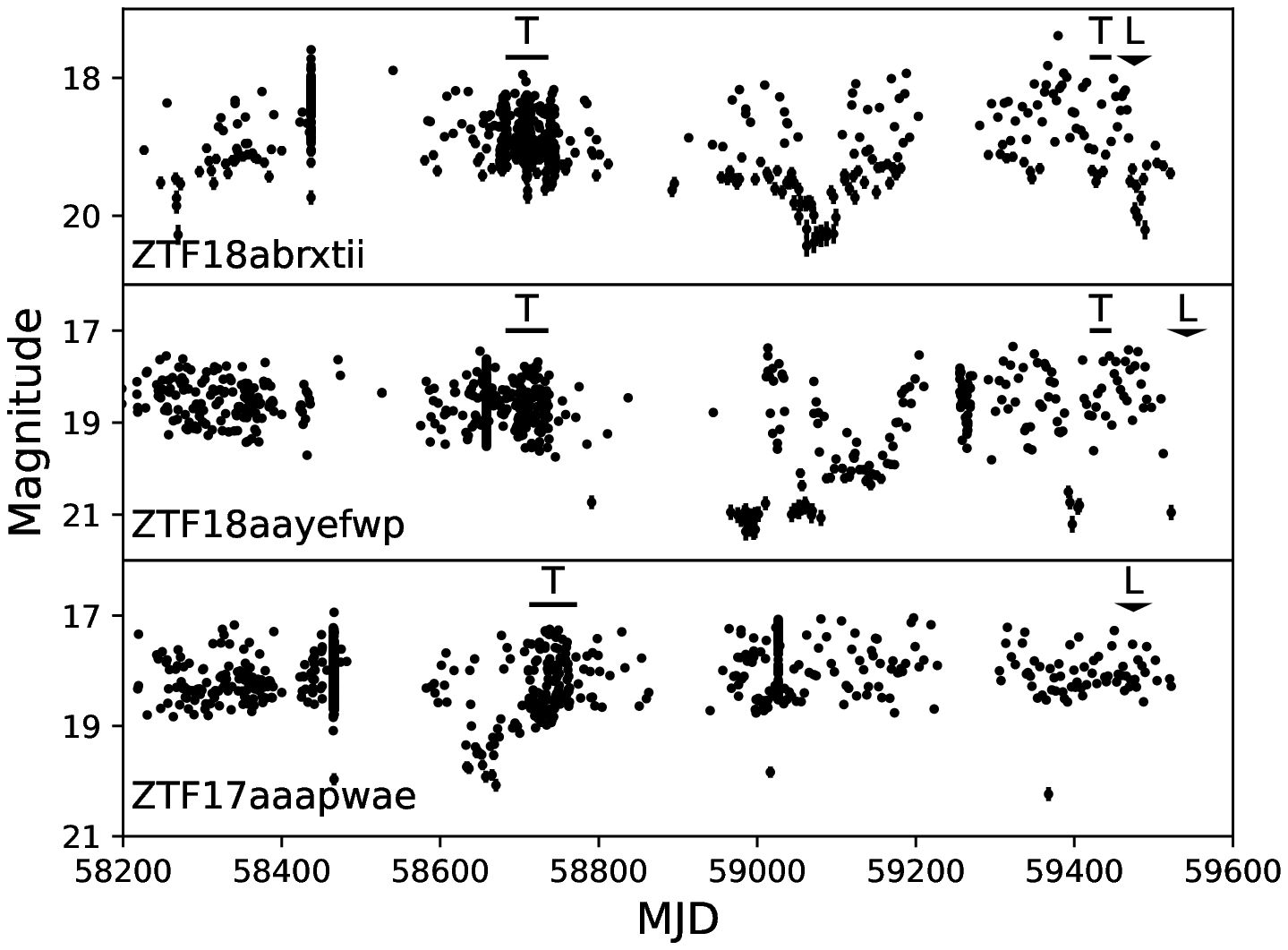}
  }
  \caption{ZTF light curves (the $r$-band filter) for G207 (ZTF~18abrxti), G205(18aayefwp)
    and G216 (17aaapwae).}
  \label{ztf3}
  \end{figure}

\begin{figure}
  \epsscale{1}
  \centerline{
    \includegraphics[scale=0.7]{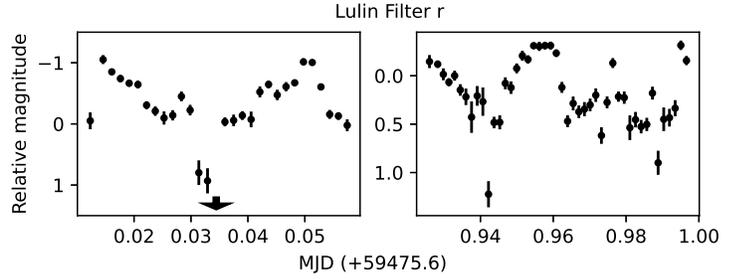}
  }
  \caption{LOT observation (the r-band filter) for G207. The observation covered two eclipses that occurred at MJD 0.034~(+59475.6) indicated by arrow symbol and  MJD 0.94~(+59475.6) }
  \label{18abr_BJD}
\end{figure}

\begin{figure}
  \epsscale{1}
  \centerline{
    \includegraphics[scale=0.6]{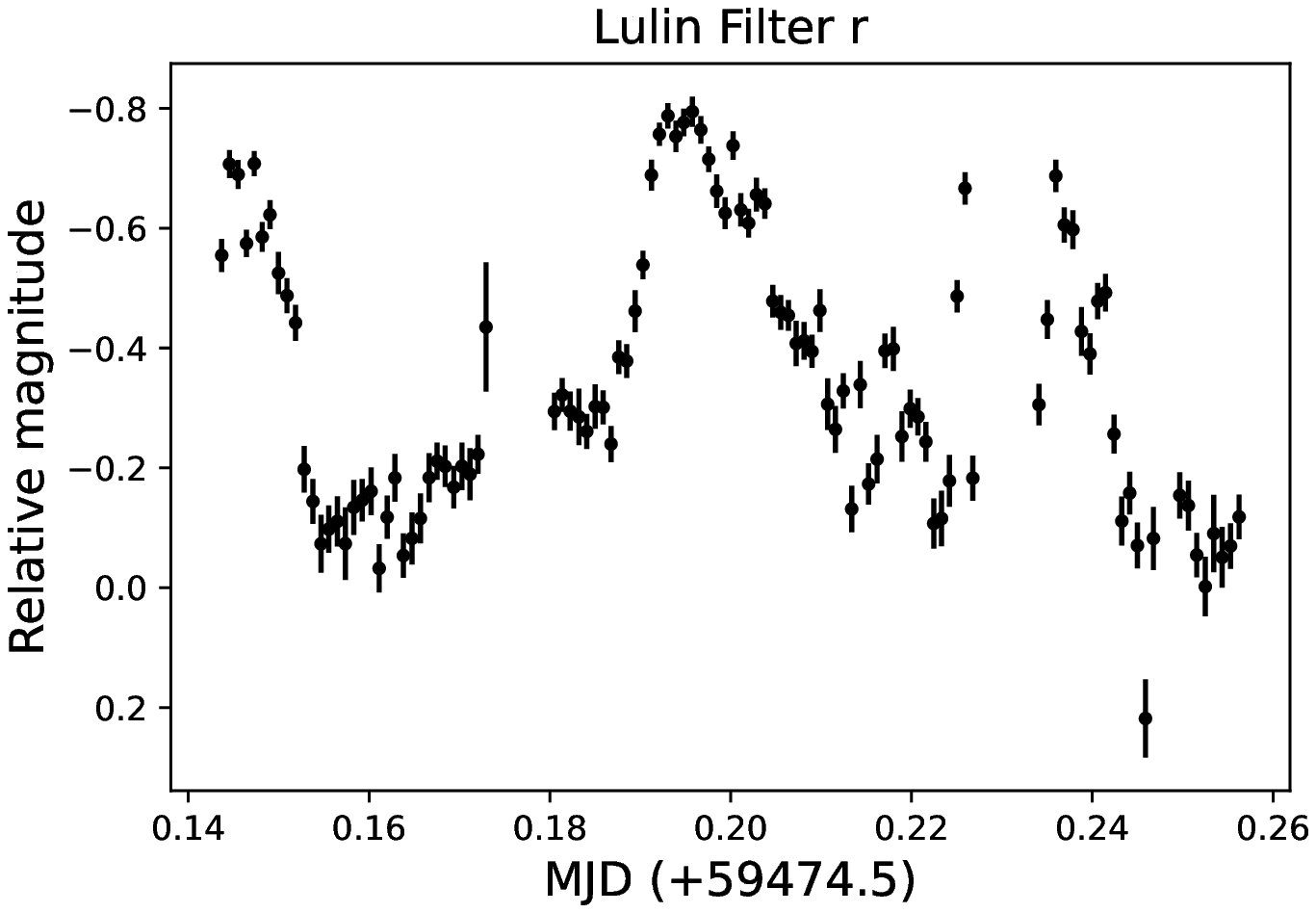}
  }
  \caption{LOT observation (the r-bands filter) for G216.}
  \label{aalight}
  \end{figure}

We find these three  candidates that do not show frequent outbursts, but the observed brightness drops  suddenly
and stays at a low luminosity state with  a time scale of months. Figure~\ref{ztf3} shows the light curves of GAIA DR2 2072080137010352768
(hereafter G207), 2056003803844470528 (G205),  and 2162478993740496256 (G216)  measured by ZTF, which are named  ZTF~18abrxtii, 18aayefwp, and 17aaapwae, respectively. For G207 (top panel of Figure~\ref{ztf3}), for example, a sudden
drop in brightness occurred   around MJD~59050 and the source  stayed in  a low luminous state for $\sim 50$~day. These three GAIA sources are selected as the optical counterparts of Swift sources. Using the distance measured by  GAIA, the X-ray luminosity in the 0.3-10~keV energy band is estimated to be  $L_X\sim 0.5^{+0.8}_{-0.3}\times 10^{31}(d/313~{\rm pc})^2~{\rm erg~s^{-1}}$ for G207, $\sim 1.4^{+3.4}_{-0.6}\times 10^{31}(d/~595{\rm pc})^2{\rm erg~s^{-1}}$ for G205 and $\sim 1.2^{+0.4}_{-0.3}\times 10^{31}(d/429~{\rm pc})^2 {\rm erg~s^{-1}}$ for G216, respectively (Table~3).

For these three targets, we can detect significant periodic signals in the ZTF and TESS  (Tables~1~and~2). The photometric periodic signals in the ZTF light curves are $f_{ZTF}=28.575(1)~{\rm day^{-1}}$ ($\sim 0.035$~day) for G207, 
$24.595(1)~{\rm day^{-1}}$ $(\sim 0.041$~day) for G205  and $22.175(1)~{\rm day^{-1}}$ ($\sim 0.045$~day) for G216, respectively. We also obtain consistent periodic signals in the TESS light curves.  

We carried out photometric observations for the three sources with LOT.
For G207 (Figure~\ref{18abr_BJD}), although the data fluctuation is significant, we find an eclipsing feature in the light curve. Figure~\ref{18abr_BJD} shows that  the source became  fainter around MJD 0.034~(+59475.6) and  MJD 0.094~(+59475.6), and the time interval between two epochs is consistent with an integer multiple of the  photometric period of $1/f_{ZTF}\sim 0.035~{\rm day}$. The orbital period, however, will be double of the photometric period, since the observation in left panel of Figure~\ref{18abr_BJD} should cover another eclipse if the orbital period is $\sim 0.035$~day. ZTF/TESS light curves may indicate a secondary eclipse with a shallower depth (Figure~\ref{all-light}).

For G216 (Figure~\ref{aalight}), the light curve indicates a  periodic modulation, and the time interval between  two strong peaks is  consistent with the photometric period of $1/f_{ZTF}\sim 0.045$~day, and the LOT light curve is consistent with the TESS/ZTF light curves (Figure~\ref{all-light}). Hence, G216 is a binary system with an orbital period of  $0.045$~day. For G205,  we could carry out only $\sim 40$~minute observation with LOT. The data, therefore, is not enough to constrain the orbital period of the source.

\subsection{GAIA DR2 4321588332240659584}
\label{g432}
\begin{figure*}
  \epsscale{1}
  \centerline{
    \includegraphics[scale=0.55]{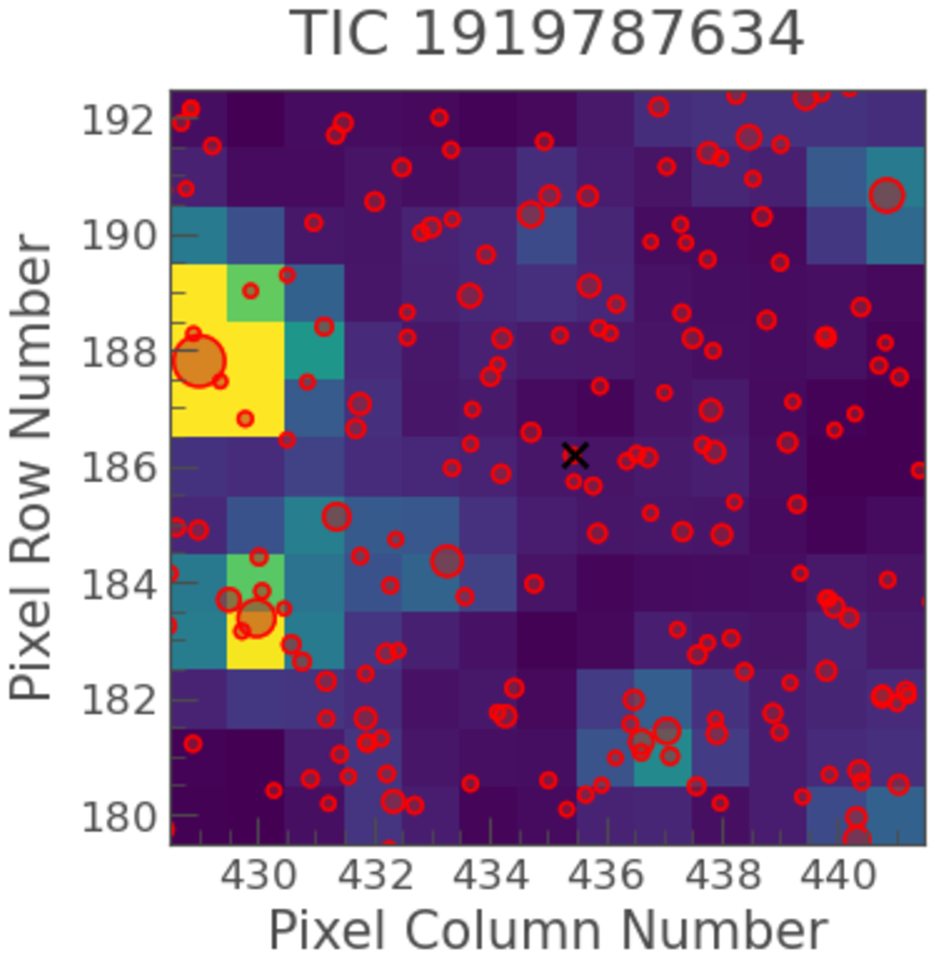}
    \includegraphics[scale=0.5]{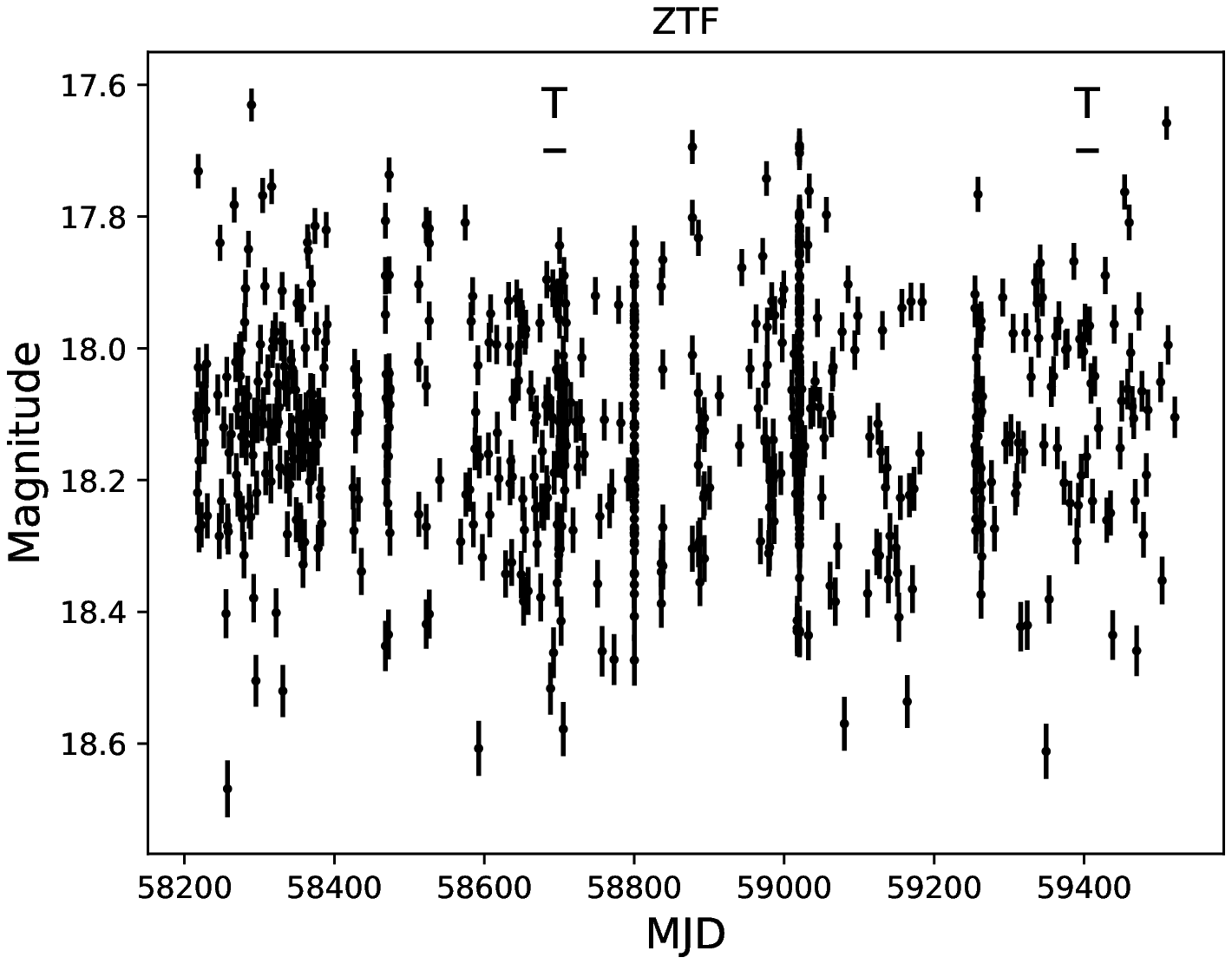}
  }
  \caption{Left: TESS-FFI image around G432, overlaid by the GAIA sources ($m_g>20$). The cross at the center of image indicates the position of the target. Right: The light curve of the target measured by ZTF (ZTF18aazmehw).}
  \label{ztf18aaz}
\end{figure*}

\begin{figure*}
  \epsscale{1}
  \centerline{
    \includegraphics[scale=1]{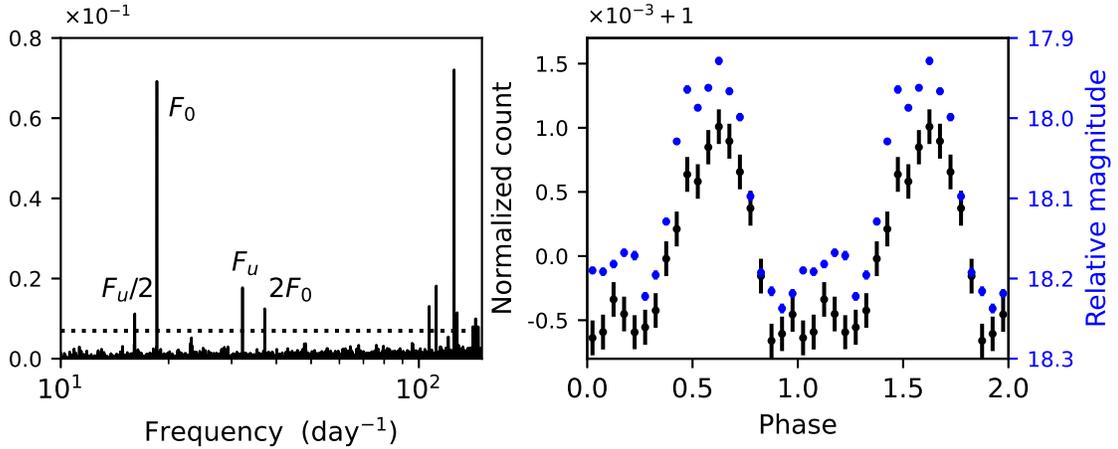}
  }
  \caption{Left; LS-periodogram for light curve of G432 measured by TESS. $F_0$ and $2F_0$ are frequencies detected in the ZTF light curve and its harmonics, respectively. $F_u$ and $F_u/2$ are unknown frequencies and its harmonics, respectively. Right: The  ZTF (blue) and TESS (black) light curve  folded with
    a frequency of $f_{ZTF}=18.552~{\rm day^{-1}}$ ($\sim 0.054$~day). Phase zero corresponds to MJD~59398.713391. The horizontal dotted line shows FAP=0.01 estimated with the bootstrap method. }
  \label{18aaz-tess}
\end{figure*}

\begin{figure*}
  \epsscale{1}
  \centerline{
    \includegraphics[scale=1]{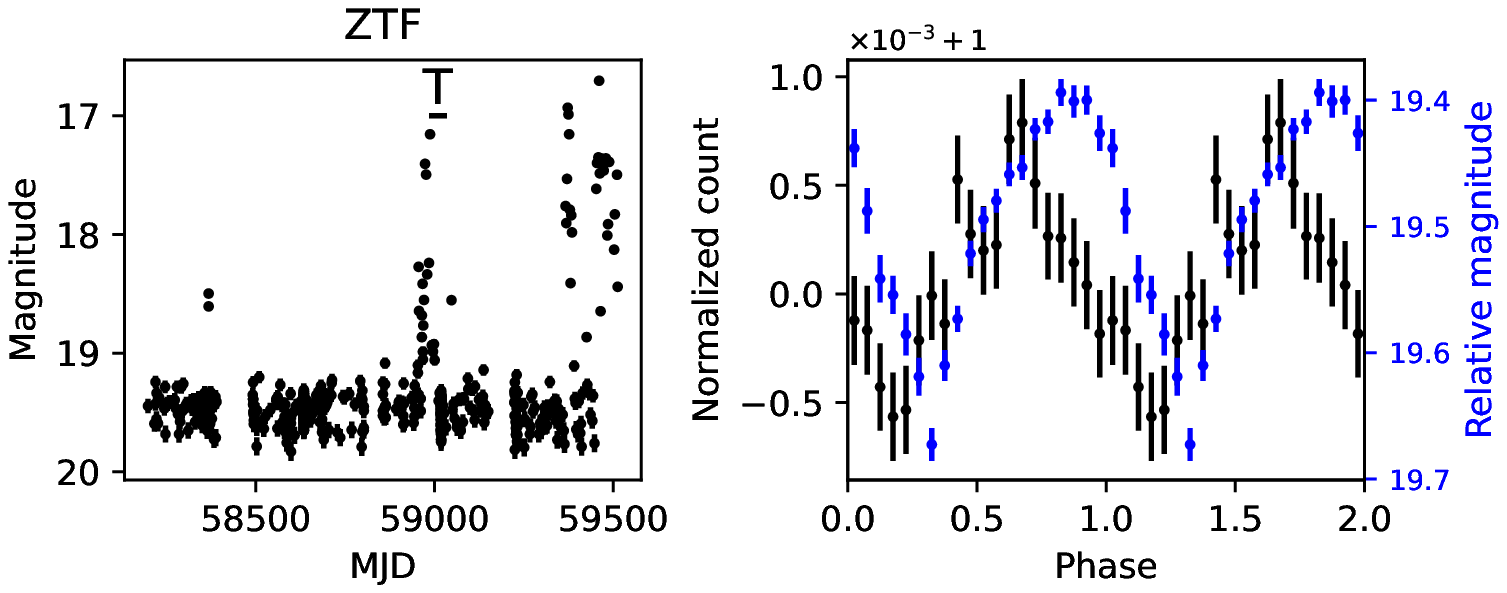}
  }
  \caption{Left:  The light curve of G454 measured  by ZTF (ZTF18abttrrr). The outburst occurred at 2021, September as reported as  a GAIA transient source, AT~2021aath (GAIA21eni). Right : ZTF (blue) and TESS (black) light curves folded with a frequency of $f_{ZTF}=8.6499~{\rm day^{-1}}$ ($\sim 0.12$~day), and the phase zero refers  MJD~59398.713391. The TESS observation was carried out
    around MJD~58980-5904 during the outburst.  }
  \label{18abt}
  \end{figure*}
GAIA DR2 4321588332240659584 (hereafter G432) is selected as a candidate for a  counterpart of the Swift point
source, 2SXPS~J192530.4+155424, and the  X-ray luminosity is measured as   
$L_X\sim  1.4^{+2.4}_{-0.8}\times 10^{32}(d/582~{\rm pc})^2{\rm erg~cm^{-2}s^{-1}}$ in the 0.3-10~keV band (Table~3). As compared with the ZTF light curves of other candidates in this study, the optical emission from this source (ZTF18aazmehw) 
is more steady with an amplitude of   variation of  less than
$\sim$1 mag (right panel of Figure~\ref{ztf18aaz}). In the LS-periodogram of the ZTF light curve, we confirm a periodic signal at  a frequency of $f_{ZTF}=18.5521(9)~{\rm day^{-1}}$ ($\sim 0.054$~day).

TESS observed the region around G432 in  2019, July (sector 14) and 2021, June (sector 40), for which data were taken approximately every $30$~minutes and $\sim10$~minutes, respectively. Figure~\ref{18aaz-tess} shows LS-periodogram  for the  data of
sectors~14 and 40. The LS-periodogram (left panel) clearly indicates a periodic signal at  $F_0=18.553(1)~{\rm day^{-1}}$, which is consistent with the finding from ZTF. The folded light curve with $f_{ZTF}$ (right panel of Figure~\ref{18aaz-tess}) 
 may be  described
by  the main peak plus a small secondary, rather than a pure sinusoidal curve, although the  significance of the secondary peak is small.
In the LS-periodogram, we  can see the signals at the second harmonics ($F_1=2F_0$) and
 the effect of aliasing with the data  sampling  frequency of $F_s\sim 145~{\rm day^{-1}}$, namely  $F_{b}=F_s-F_0\sim 126~{\rm day^{-1}}$ and  $F_{b1}=F_s-F_1\sim 108~{\rm day^{-1}}$

In addition to the periodic signals related to $F_0$ and $2F_0$, the periodogram shows other signals at
$F_u=32.17(4)~{\rm day^{-1}}$ ($\sim 0.031$~day) and its harmonics $F_u/2$. This signal clearly appears 
in the TESS  data of the sector~40. To check which pixel of the TESS-FFIs causes the  signal,
we extract the light curve of each pixel around the sources for the observation of sector~40. We find that a significant  signal with $F_u$
can be detected  at one pixel where the power of the signal with $F_0$ becomes the maximum.
The periodic signal with $F_0$ can be seen in other pixels but  the power of the signal is lower than  that at  the pixel where the $F_u$ signal is found. Hence, the signal, $F_u$,  may be  related to our target, although other optical sources located
in the same pixel (right panel of Figure~\ref{ztf18aaz}) cannot be ruled out as the origin of the signal. 
  We note that LS periodogram of TESS data is insensitive to the time-correlated noise model discussed in Appendix~\ref{noise}.

If the signal $F_u$ is  related to  G432, then the origin may be related to  the harmonics of $F_0$.
Although the frequency $F_u=32.17(4)~{\rm day^{-1}}$ cannot be described  by a simple relation with $F_0$,  the frequency $F_s-F_u\sim 6F_0$,
where $F_s\sim 145~{\rm day^{-1}}$ is the data readout frequency, indicates that the signal $F_u$  is related to  the 6th harmonic of the fundamental frequency. It is not trivial,  however, why the modulation of the 6th harmonic is more evident than those of the 2nd-5th harmonics. 
 If the frequency  $F_u$ is independent of $F_0$,  
 the signal would  be related to the spin of the WD.  We cannot find a significant periodic signal at the frequency  $F_u$ in the data of  sector~14.  This  may be due to a large uncertainty of each data point of the observation or the true signal is 
$F_s-F_u\sim 111~{\rm day^{-1}}$ ($\sim 13$~minutes)  for which the data of sector~14 cannot identify. If $F_u$ and $F_0$ are related to the WD spin and orbital period, respectively, G432 could  be a candidate for IP due to the fact that the X-ray emission dominates the UV emission, which is a typical property of the emission of  IPs (section~\ref{comp}). A data set taken with a higher cadence is required to identify the origin of $F_u$.

\subsection{GAIA DR2 4542123181914763648}
\label{g452}
We select GAIA DR2~4542123181914763648 (hereafter G452)  as a candidate of the optical  counterpart of
1RXS~J172728.8+132601, for which a 2.5~ks Swift observation measures the X-ray luminosity of $L_X=5.5^{+14}_{-3.2}\times 10^{31}(d/502{\rm pc})^2~{\rm erg~s^{-1}}$ with $N_H=7.5\times 10^{21}~{\rm cm^{2}}$ (Table~3).  As the ZTF light curve (ZTF18abttrrr) shows (left panel of Figure~\ref{18abt}), the source repeats outbursts on  a time scale of years; the outburst happened in 2021 was alerted as  a GAIA transient source,
AT~2021aath \citep[GAIA21eni,][]{2021TNSTR3431....1H}, which was identified as a polar type CV. Nevertheless,
since there is no  detailed binary  information for the target in the literature, we search for a possible periodic signal in the ZTF light curve.
  After removing the data of the outburst from the light curve, the  LS-periodogram shows a significant periodic signal at    $f_{ZTF}=8.6499(8)~{\rm day^{-1}}$ ($\sim 0.115$~day).

  TESS observation for the region  around G452   was carried out in 2020,
 May and June (sectors 25 and 26)  during the outburst around MJD~59000 (Figure~\ref{18abt}). 
 We extracted the  TESS-FFI light curve and removed  a long-term
 trend  caused by the outburst from  the light curve. We found a periodic signal with $F_0=8.65(2)~{\rm day^{-1}}$, which is consistent with the result of the ZTF observation. The right panel of Figure~\ref{18abt} shows the folded light curves of the ZTF and TESS data, which can be described by a sinusoidal modulation. As Figure~\ref{18abt} shows, we also observe a shift in the  TESS light curve from the ZTF light curve, which may be due to the effect of an outburst during the TESS observation.  We do not find another  significant periodic signal in the TESS data, which may be consistent  with a spin-orbit phase synchronization of the polar.

 \subsection{Other candidates}
\label{others}
 We searched for a possible periodic signal of  29 GAIA DR2  sources, which are potential counterparts of  an unidentified X-ray source.  Table~4 presents four  sources, for which  TESS-FFI data provides a periodic
 signal,  but no enough  data points  or no data are available for ZTF observations. As we have mentioned, there are two observational modes  for TESS-FFIs,
 in which the data readout time intervals are $\sim48~{\rm day^{-1}}$ and $\sim144~{\rm day^{-1}}$, respectively. As shown in  Table~4, GAIA DR2 5360633963010856448 and 5964753754945126528 were observed by two observational modes, and the periodic signal $F_0$ reported in Table~4  is identified in both data sets. For GAIA~DR2~2031371479242995584 and~ 1981213682883140864, on the other hand, data from only one observational mode  is available, and we  cannot discriminate between the  true  frequency and the aliasing. We note that due to one pixel of the TESS observations containing several GAIA sources, we cannot rule out that the periodic signal is related to another source. 

 Among the four sources in Table~4,  GAIA DR2 5964753754945126528 (hereafter G596), which shows a periodic signal $F_0\sim 8.34~{\rm day^{-1}}$ $(\sim 0.120$~day) in the TESS data,  is a promising candidate for a CV, and its location is consistent  with
 the X-ray source,  AX~J1654.3-4337, which
 was discovered by the ASCA Galactic plane survey~\citep{2001ApJS..134...77S}.    Swift  and NuSTAR  observations for this source were carried out in 2020 July-August  with a total exposure of 6.7~ks and 26~ks, respectively.  We extract the events  and spectra of the source region
 with the command \verb|nupipeline|   under \verb|HEASoft ver.6-29| for the NuSTAR data and the package \verb|Xselect| for the Swift data.

\begin{figure}
  \epsscale{1}
  \centerline{
    \includegraphics[scale=0.55]{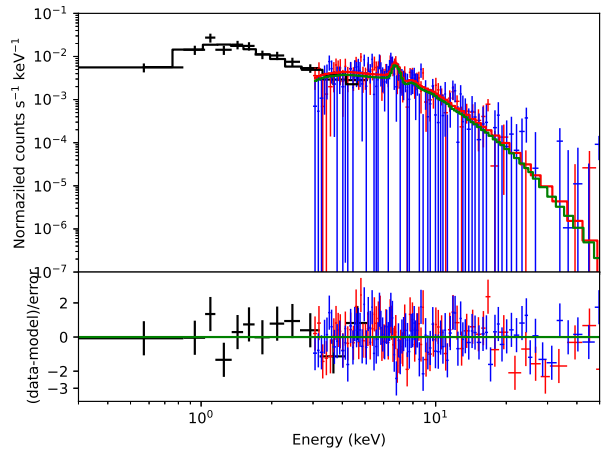}}
  
  \caption{Swift (black) and NuSTAR (red and blue) spectra of AX~J1654.3-4337, which will be the X-ray counterpart of GAIA DR2 5964753754945126528. The spectrum is described by an optically thin thermal plasma emission with a temperature of $k_BT\sim 8.9$~keV. }
  \label{axj16-spe}
\end{figure}

 We group the spectral bins to contain a minimum of 20~counts in each bin,  and  fit the spectrum  using \verb|Xspec ver.12.12| (Figure~\ref{axj16-spe}). We find that the spectrum in the 0.2-70~keV band is well fitted  by the optically thin thermal plasma  emission (\verb|tbabs*mekal| model in Xspec) with a temperature of $k_BT=8.9^{+2.0}_{-1.5}$~keV and absorption column density of  $N_H=2.2^{+0.08}_{-0.07}\times 10^{21}~{\rm cm^{-2}}$ ($\chi^2=164$  for 188 D.O.F.). The luminosity in the 0.2-70~keV band is estimated to be $L_X=6.0^{+0.6}_{-0.5} \times 10^{31}(d/462~{\rm pc})^2~{\rm erg~s^{-1}}$. The X-ray emission with a plasma temperature of $\sim 10$~keV suggests the emission from an accretion column on the WD surface, indicating  a magnetic CV system.  Moreover, the observed  X-ray luminosity of $<10^{32}~{\rm erg~s^{-1}}$ indicates that the source is not a typical  IP, for which the X-ray luminosity is typically  $>10^{32}~{\rm erg~s^{-1}}$.

\begin{figure}
  \epsscale{1}
  \centerline{
    \includegraphics[scale=0.5]{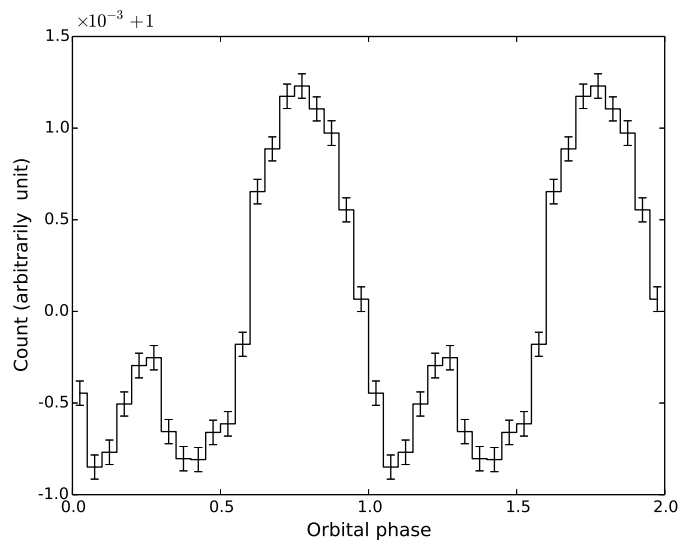}}

  \caption{TESS light curve of G596/AX~J1654.3-4337 folded with $F_0=8.34~{\rm day^{-1}}$ $(\sim 0.12~{\rm day})$. The phase zero refers MJD~59398.713391.}
  \label{axj-tess}
\end{figure}

Figure~\ref{axj-tess} presents the folded light curve
 of $F_0=8.34~{\rm day^{-1}}$ extracted from the TESS-FFI data. 
We can see that the shape of light curve shows  a double-peak structure.
Combined with the X-ray spectral properties, this optical emission likely originated from  two  magnetic poles 
heated by the accretion column, and the modulation is likely due to the WD spin. Since the TESS data did not indicate other periodic signals longer than this WD's spin signal within 10~day, 
G596/AX~J1654.3-4337 may be a polar, although we cannot rule out the possibility of an IP.  We do not find any periodic signal in the NuSTAR data, in which  a window effect of the observation dominates the LS-periodogram.

\begin{deluxetable*}{cccccc}
\tablecolumns{5}
\tabletypesize{\footnotesize}
\tablecaption{Other  CV candidates}
\tablehead{
\colhead{GAIA}  & 
\colhead{X-ray source} &
\colhead{Distance}  &
\colhead{TESS}  &        
\colhead{$F_0$\tablenotemark} &
\colhead{Proposed type} \\
\colhead{DR2} &
\colhead{}& 
\colhead{(pc)}  &
\colhead{Sector}  &      
\colhead{(${\rm day^{-1}}$)} &
\colhead{}
 }
\startdata
    5360633963010856448    &   1RXS J104612.9-511819 & 496 & 10, 36, 37 & 15.46(2)\tablenotemark{\rm a}  &  \\
    5964753754945126528   &    2SXPS J165423.6-433745 & 462 & 12, 39 & 8.34(4)\tablenotemark{\rm a} & Polar \\
     & AX J1654.3-4337 (AX1654) &  & & \\
    2031371479242995584 & 1RXS J194401.5+284456  & 416 & 40, 41 & 4.20/139.8\tablenotemark{\rm b}  & \\
    1981213682883140864 & 2SXPS J220344.5+525450 & 754 & 16, 17  &1.15/46.85\tablenotemark{\rm b} &  
\enddata
\tablenotetext{\rm a}{The periodic signal is identified in two different modes}
\tablenotetext{\rm b}{Only data taken by one mode is available.}
\end{deluxetable*}

\section{Discussion and summary}
\label{discuss}
\subsection{Comparison with known CVs}
\label{comp}
\begin{figure}
  \epsscale{1}
  \centerline{
    \includegraphics[scale=1]{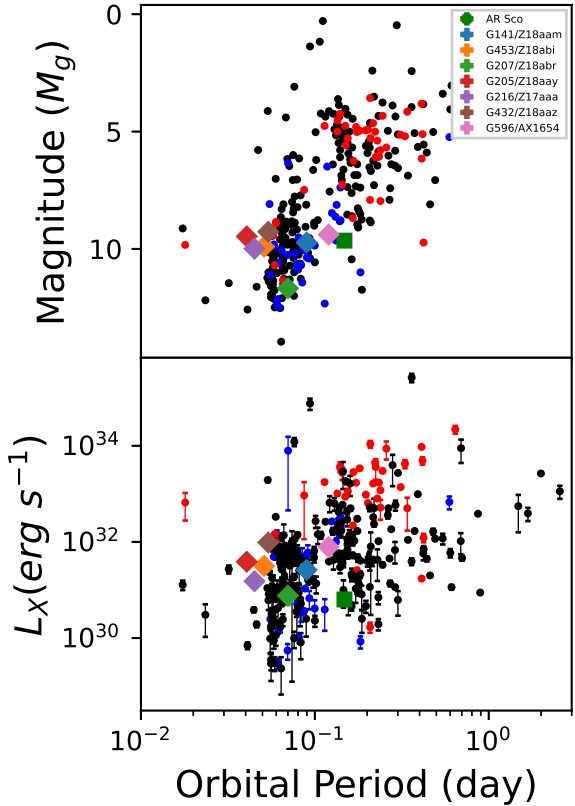}}
  \caption{Distributions of GAIA G-band magnitudes (top) or X-ray luminosity in the 0.3-10~keV band with the orbital period: For G453, G205 and G432, the orbital period is assumed to be $F_0$ measured by TESS, although true period of  $F_0/2$ cannot be ruled out. The back, blue and red filled circles correspond to nonmagnetic CV, polar and IP, respectively,  as listed in Ritter Cataclysmic Binaries Catalog \citep{2003A&A...404..301R}. The diamond  and green box symbols represent CV candidates 
and AR Scorpii, respectively. The X-ray luminosity is calculated from the flux and parallax 
measured by Swift \citep{2020ApJS..247...54E} and GAIA, respectively.  }
  \label{dis}
\end{figure}

\begin{figure}
  \epsscale{1}
  \centerline{
    \includegraphics[scale=0.55]{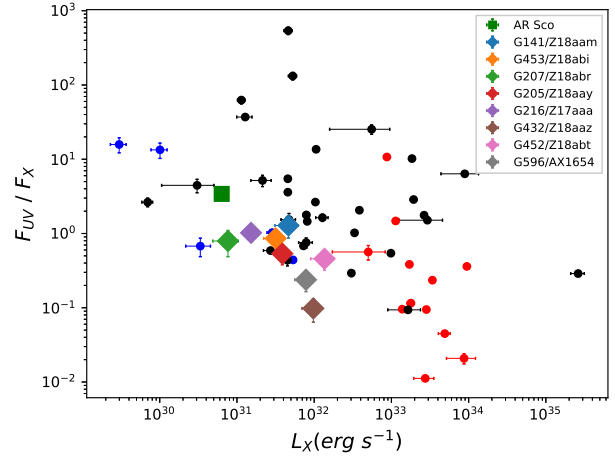}}
  \caption{Distribution of UV/X-ray flux ratio with the X-ray luminosity measured by the Swift observation. The back-, blue-, and red-filled circles correspond to no-magnetic CV, polar, and IP, respectively. }
  \label{lxuvx}
\end{figure}

Figures~\ref{dis} and~\ref{lxuvx} compare the  properties of our CV candidates with those of known CVs. Figure~\ref{dis} shows the distributions of  the orbital periods and GAIA G-bands magnitudes (upper panel) or the 
X-ray luminosity in the 0.3-10~keV band (lower panel) of our candidates. In the figure, we can see  so-called period gap at $P_{orb}\sim 2-3$~hr, in which less CVs have been detected \citep{1983A&A...124..267S, 2010MmSAI..81..849R, 2003Ap.....46..114K,2018ApJ...868...60G}.  The figure shows that the nonmagnetic systems (black-filled circle) have been detected with a period in the range of 0.01-1~day. Known IPs (red-filled circle) and  polars (blue filled circle), on the other hand,
concentrate  on the orbital periods  longer and shorter than the period gap, respectively.  We find in the figure  that 
the six  CV candidates discussed in the current study  are located  below the 
 period of the gap.  This will be a selection effect because (i) there is a correlation between 
the orbital period and GAIA G-band magnitudes, as seen in top panel of Figure~\ref{dis}, and (ii) we have selected the candidates with the condition  $9<M_G<12$ of the G-bands magnitude  (Figure~\ref{hr}).
AX J1654.3-4337 (pink diamond), which is a  polar candidate  as discussed in section~\ref{others},   
 locates in or close to the period gap.  As the bottom panel of Figure~\ref{dis} shows, the X-ray luminosity of our candidate, $L_X\sim 10^{31-32}~{\rm erg~s^{-1}}$ is relatively larger among the CVs located below the period gap.   This is  because we searched   candidates in the  X-ray source catalogs of previous surveys, which  may have  miss  many faint X-ray sources.  

 We can see in the figure  that four candidates (G453, G205, G216, and G432) have an orbital period close to or shorter than $\sim 0.05$~day, which is known as the  minimum orbital period in the standard binary evolution of 
 CVs \citep{2010MmSAI..81..849R}. We note  however that the orbital periods of G453, G205 and G432 may be double  the values in the figure (sections~\ref{threes} and~\ref{g432}), and hence they may have an orbital period longer than the minimum period.

Figure~\ref{lxuvx} shows the distribution of the 
flux ratio of the UV band and X-ray band measured by Swift.  As can be seen in the figure,  the ratio is greater than unity  for 
most of the  nonmagnetic system, which is understood by the emission from the boundary layer of the accretion disk. 
For magnetic CVs, on the other hand, the X-ray dominates the UV emission, which is  a characteristic of emission from the accretion column~\citep{2017PASP..129f2001M}. Several sources, on the other hand,  have a 
ratio greater than 10, and the emission 
is  probably dominated by  blackbody emission from the pole with a temperature of $<50$~eV.

We find that our candidates have relatively hard spectrum ($F_{UV}/F_X\le 1$) and the  two  hardest sources   (G432 and G596) are  classified as candidates for  magnetic CVs. For G432, we identify 
two possible periodic signals ($F_0=18.5~{\rm day^{-1}}$ and $F_u=32.2$ or $111~{\rm day^{-1}}$) in the TESS data. With the UV/X flux ratio much smaller than unity, G432 may be a candidate of X-ray faint  IP. For G596, the properties of the optical light curve and the X-ray spectrum are consistent with a polar, as described in section~\ref{others}.

In the unidentified X-ray source, we searched for new binary 
candidates that show emission properties similar to those of  AR~Scorpii, for  which (i) no mass is probably transferred from the companion star to the WD, (ii) UV emission dominates 
the X-ray emission, and (iii) a magnetic WD heats up the companion star.  With current photometric studies, however, we can conclude that these candidates do not belong to the  AR Scorpii-type binary system. For example, three candidates (G141, G453 and G454) clearly show 
dwarf-nova-type outbursts, suggesting the existence of a mass transfer from the companion to the WD, and thus an accreting system.
With a large optical variation ($\delta m\sim 1.5$), AR Scorpii contains a WD heating up 
the companion star. The optical curves of our candidates show   a smaller  magnitude variation ($\delta m< 1$, e.g. Figure~\ref{aalight} for G216 and Figure~\ref{ztf18aaz} for G432), and do not show 
 any clear evidence  of heating.  
 G596/AX~J1654.3-4337 is a   promising candidate for the magnetic system, but it is a polar, while AR Scorpii is an IP in the sense that the spinning period is different from the orbital period.

 Finally, we note that out of  our  eight candidates,  six systems   have not experienced a large outburst in the last 4 years, and  have been missed in the previous surveys. Our study, therefore,  will  suggest
 that a large population of  WD binary systems with an inactive mass transfer could be discovered in future surveys.

In summary, we searched for new CV systems associated with the unidentified X-ray sources listed in ROSAT, Swift and XMM-Newton source catalogs. We selected  the GAIA sources with a g-band magnitude of $9<M_G<12$ and a color $0.5<G_{BP}-G_{RP}<1.5$, 
and identified 29 sources that are potential counterparts of the unidentified X-ray sources.
 We carried out a  photometric study with  
 ZTF, TESS and LOT observations, and  we constrained the orbital periods for the seven sources (sections~\ref{g141}-\ref{g452}). Among the seven candidates, G141 and  G207 are  eclipsing binary systems, and G141 shows a secondary eclipse with a shallower depth. We identified 
 three candidates (G141, G453 and G454), in which 
 a mass transfer from the companion to WD 
is active, and they   exhibit repeated 
 outbursts (Figures~\ref{ztf18aam}, \ref{ztf18abi} and \ref{18abt}). For the other three 
candidates (G207, G205, and G216), on the other hand, it is observed that    
the source brightness suddenly drops and stays at a low luminous state (Figure~\ref{ztf3}), suggesting 
the mass transfer may be inactive. Based on the detection of two periodic signals in the light curve and UV/X-ray emission properties, we  can classify G141 and G432 as IP candidates, although a spectroscopic study is required for confirmation. In addition to seven candidates, we confirmed that the  unidentified ASKA source, AX J1654.3-4337 (G569), is a candidate for being  a polar.
With the current photometric studies, we could not find any candidate for the AR Scorpii-type WD binary, although we selected the GAIA sources that had a similar
  magnitude  to AR Scorpii.   AR Scorpii still remains  a n 
unique WD binary system, and a more sophisticated search (e.g. an optical survey for unidentified X-ray source with high cadence)  will be required to 
find a new AR~Scorpii-type binary system.

\vspace{4mm}
We thank to referee for his/her useful comments and suggestions.
We are grateful to  Dr. Kato for providing the source list in VSX catalog
 and sending some  references.   
We also thank  the Swift-TOO team for arranging the  observations for our sources. J.T. and X.F.W are supported by
the National Key Research and Development Program of China (grant No. 2020YFC2201400) and the National Natural Science Foundation of China (grant No. 12173014). A.K.H.K. is supported by the Ministry of Science and Technology (MOST) of Taiwan through grant Nos. 108-2628-M-007- 005-RSP and 109-2628-M-007-005-RSP. J.M. is supported by the National
Natural Science Foundation of China (grant No. 11673062). C.-P.H. acknowledges support from the MOST of Taiwan through grant MOST 109-2112-M-018-009-MY3. L.C.-C.L. is supported by  MOST through grant MOST 110-2811-M-006-515 and MOST 110-2112-M-006-006-MY3. K.-L. L. is supported by the MOST of Taiwan through grant 110-2636-M-006-013, and he is
a Yushan (Young) Scholar of the Ministry of Education of Taiwan. C. Y. H. is supported by the National Research Foundation of Korea through grant 2016R1A5A1013277 and 2019R1F1A1062071.

{\it Note added in proof:}
While this article in press, we were noticed that potential orbital 
periods of 4 sources have been already reported in the  AAVSO catalog (for 2SXPS J202600.8+333940 and 2SXPS J192530.4+155424) or in vsnet-chat (for 4XMM J172959.0+522948 and 2SXPS J195230.9+372016). We list references of the AAVSO catalogs and archives of vsnet-chat for the 7 sources listed in Table~1. In Appendix C (for manuscript in archive), we briefly mention the information in the catalog and compare with our results.  

{\tt ZTF18aampffv=4XMM J172959.0+52294=MGAB-V705}
\url{https://www.aavso.org/vsx/index.php?view}\\
\url{=detail.top&oid=1499054} \\
\url{http://ooruri.kusastro.kyoto-u.ac.jp/mailarchive/}\\
\url{vsnet-chat/8923}

{\tt ZTF18abikbmj=1RXS J185013.9+242222=DDE163}
\url{https://www.aavso.org/vsx/index.php?view=detail.top&oid=686692}

{\tt ZTF18abrxtii=2SXPS J195230.9+372016} \\
\url{https://www.aavso.org/vsx/index.php?view=detail.top&oid=2224634} \\
\url{http://ooruri.kusastro.kyoto-u.ac.jp/mailarchive/vsnet-chat/8866}

{\tt ZTF18aayefwp=2SXPS J202600.8+333940=BMAM-V634}
\url{https://www.aavso.org/vsx/index.php?view=detail.top&oid=1543125}\\
\url{http://ooruri.kusastro.kyoto-u.ac.jp/mailarchive/vsnet-chat/8920}

{\tt ZTF17aaapwae=2SXPS J211129.4+445923} \\
\url{https://www.aavso.org/vsx/index.php?view=detail.top&oid=2223038}

{\tt ZTF18aazmehw=2SXPS J192530.4+155424=DDE182}
\url{https://www.aavso.org/vsx/index.php?view=detail.top&oid=1543028}

{\tt ZTF18abttrrr=1RXS J172728.8+132601=GAIA21eni}\\
\url{https://www.aavso.org/vsx/index.php?view=detail.top&oid=2224470}

\facility{{\it Swift}(XRT), {\it XMM-Newton}(EPIC), {\it NuSTAR}(FPM), {\it ZTF}, {\it TESS} and {\it LOT}}.

\software{\newline {\tt Science Analysis System} \\ (\url{https://www.cosmos.esa.int/web/xmm-newton/how-to-use-sas}; \citealt{SAS2004})
  \newline {\tt HEASoft} \\ (\url{https://heasarc.gsfc.nasa.gov/docs/software/lheasoft/\\developers\_guide/}; \citealt{HEAsoft2014})
  \newline {\tt Xspec} \\ (\url{https://heasarc.gsfc.nasa.gov/xanadu/xspec/}; \citealt{Xspec96})
  \newline {\tt Lightcurve} \\ (\url{https://heasarc.gsfc.nasa.gov/docs/tess/LightCurve-object-Tutorial.html}; \citealt{2018ascl.soft12013L})
  \newline{\tt eleanor} \\(\url{https://eleanor.readthedocs.io/en/latest/}\citealt{2019PASP..131i4502F})
  \newline {\tt IRAF} \\ (\url{https://iraf-community.github.io}; \citealt{1993ASPC...52..173T})
  \newline {\tt astropy} \\ (\url{https://docs.astropy.org/en/stable/index.html}; \citealt{astropy:2013})
}

\bibliography{adssample}
\appendix
\restartappendixnumbering
\section{LS periodogram with time-correlated noise model}
\label{noise}

We apply the time-correlated noise model based on Equation (13) of \cite{2020A&A...635A..83D}, and assume the covariance matrix to follow
\begin{equation}
  C_{i,j}=\delta_{i,j}\sigma_i^2+\sigma^2_{\rm exp}e^{|t_i-t_j|/\tau_{\rm exp}}, 
\label{cmodel}
\end{equation}
where $\sigma_i$ is the diagonal matrix with the observational error bars that are usually provided in the processed data,   and $\sigma_{\rm exp}$ corresponds to the correlated noise with a time scale of $\tau_{exp}$. We choose the  amplitude of the fluctuation in the light curve to be  $\sigma_{\rm exp}$, and create the periodogram for the different values of $\tau_{\rm exp}$.
For each frequency, the normalized power of the LS periodogram is calculated from  \citep{2018ApJS..236...16V} 

\begin{equation}
z_1(f)=1-\frac{\hat{\chi}^2(f)}{\hat{\chi}^2_0},  
\end{equation}
where $\hat{\chi}^2$ is the minimum value of $\chi^2(f)=(\mathbf{y}-\mathbf{y}_{\rm model})^T \mathbf{C}^{-1}(\mathbf{y}-\mathbf{y}_{\rm model})$, and  $\mathbf{y}$ and $\mathbf{y}_{\rm model}$ are the time series of the observation and model, respectively. In addition, $\hat{\chi}^2_0$ is a non-varying reference value. For each periodogram, we scan  the frequency between $1~{\rm day^{-1}}<f<100~{\rm day^{-1}}$ and estimate FAP using the method of \cite{2020A&A...635A..83D}. We can see that the LS periodogram of TESS data is insensitive to the time-correlated noise model within the current framework. 

\begin{figure*}
  \epsscale{1}
  \centerline{
    \includegraphics[scale=0.6]{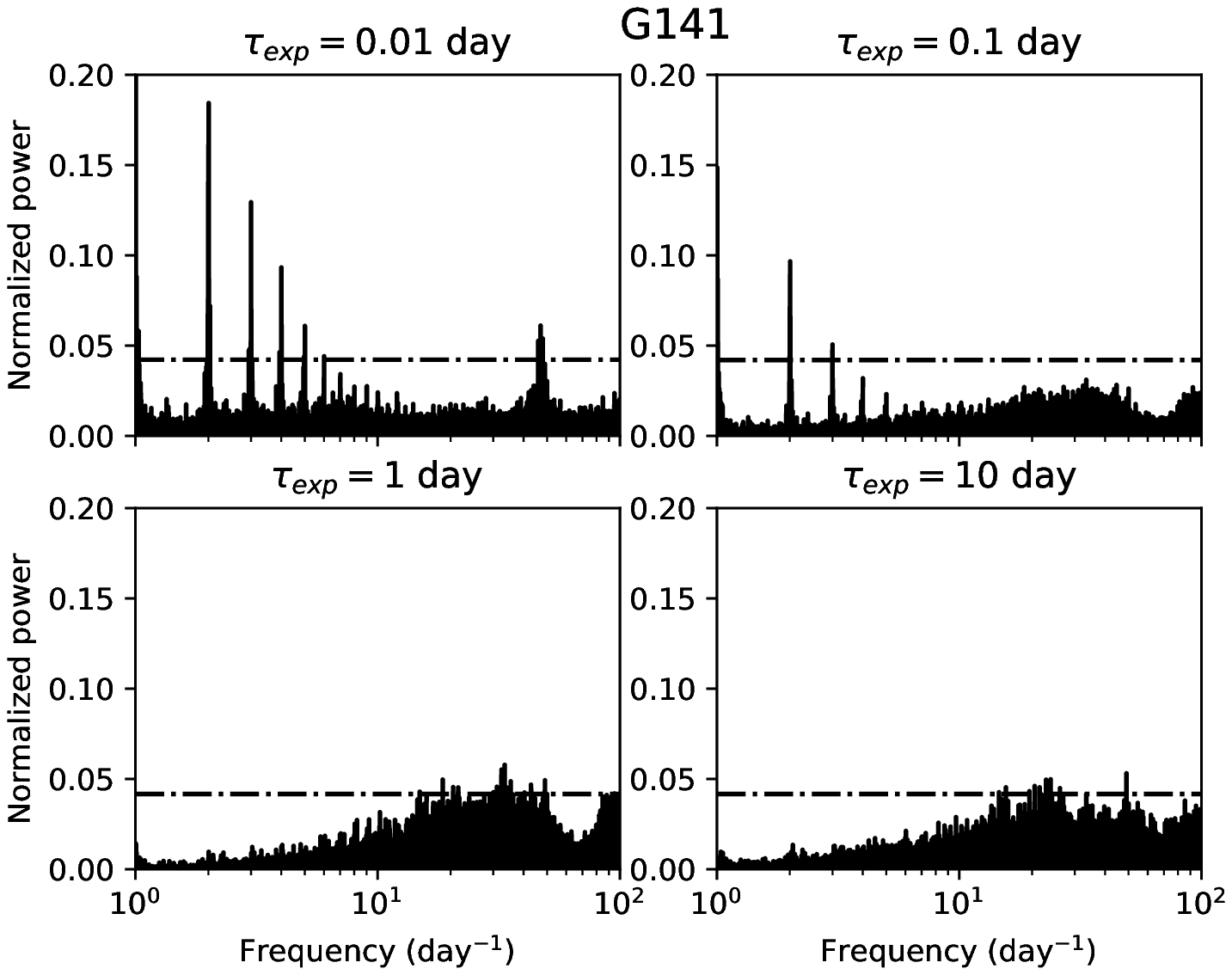}
 \includegraphics[scale=0.6]{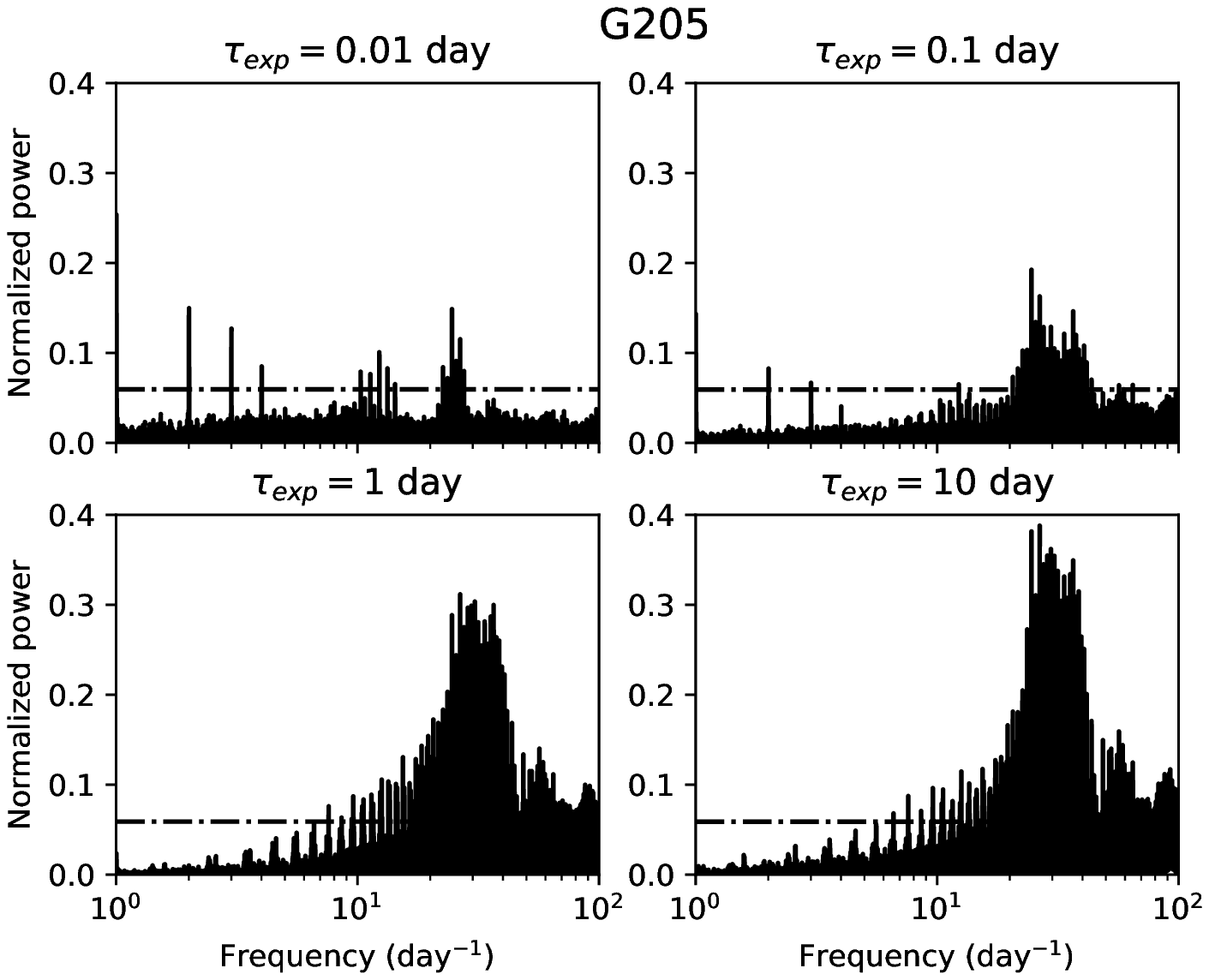}}
  
  \caption{LS periodograms with correlated noise models (\ref{cmodel}) for G141 (left) and
      G205 (right). The results are for $\sigma_{exp}=1$~magnitude. The dashed-dotted lines represent FAT=0.01 estimated from the method of \cite{2020A&A...635A..83D}. }
  \label{corre}
\end{figure*}

Figure~\ref{corre} shows the LS periodogram of the ZTF data  with the time-correlated noise model for G141 (left panel) and G205 (right panel). As we demonstrated in section~\ref{g141}, the possible periodic signal, $f_{ZFT}\sim 47~{\rm day^{-1}}$, of G141  cannot be confirmed in the TESS/LOT data. From the left panel of Figure~\ref{corre}, we find that the signal at $f_{ZFT}\sim 47~{\rm day^{-1}}$ disappears from LS~periodogram for $\tau_{exp}>0.1$~day. For G205 (right panel), on the other hand, the periodic signal $f_{ZTF}\sim 24.6~{\rm day^{-1}}$ is confirmed in the TESS data, and  the existence of the signal in the periodogram  is insensitive to the noise model.
Although the noise model affects the shape of the LS periodogram of the ZTF data, it
is less effective in determining  the existence of the periodic signal reported in Table~1.  These results suggest  that the periodic signal
$f_{ZFT}\sim 47~{\rm day^{-1}}$ of G141 is likely related to the time-correlated noise.

\begin{figure*}
  \epsscale{1}
  \centerline{
    \includegraphics[scale=0.5]{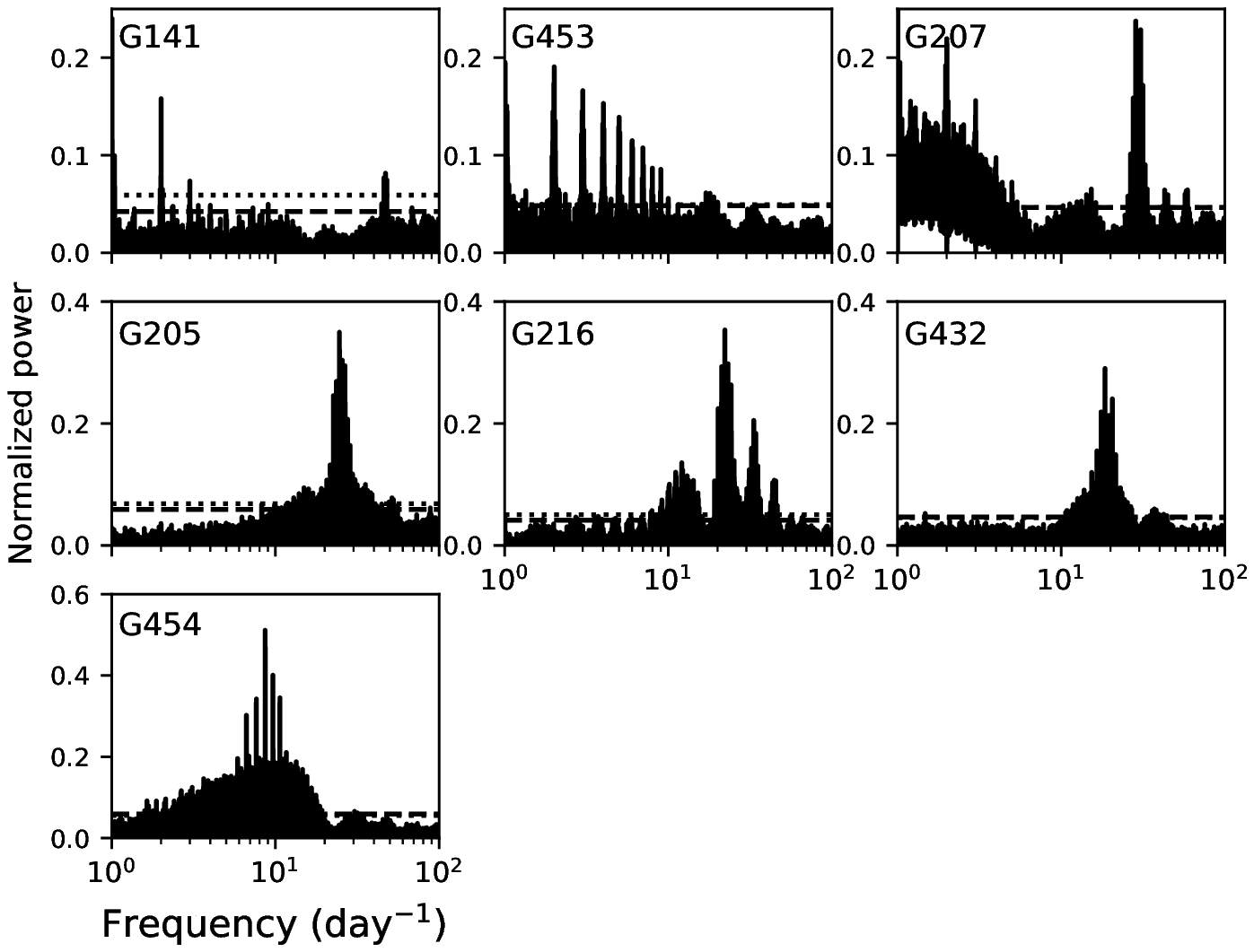}
    \includegraphics[scale=0.5]{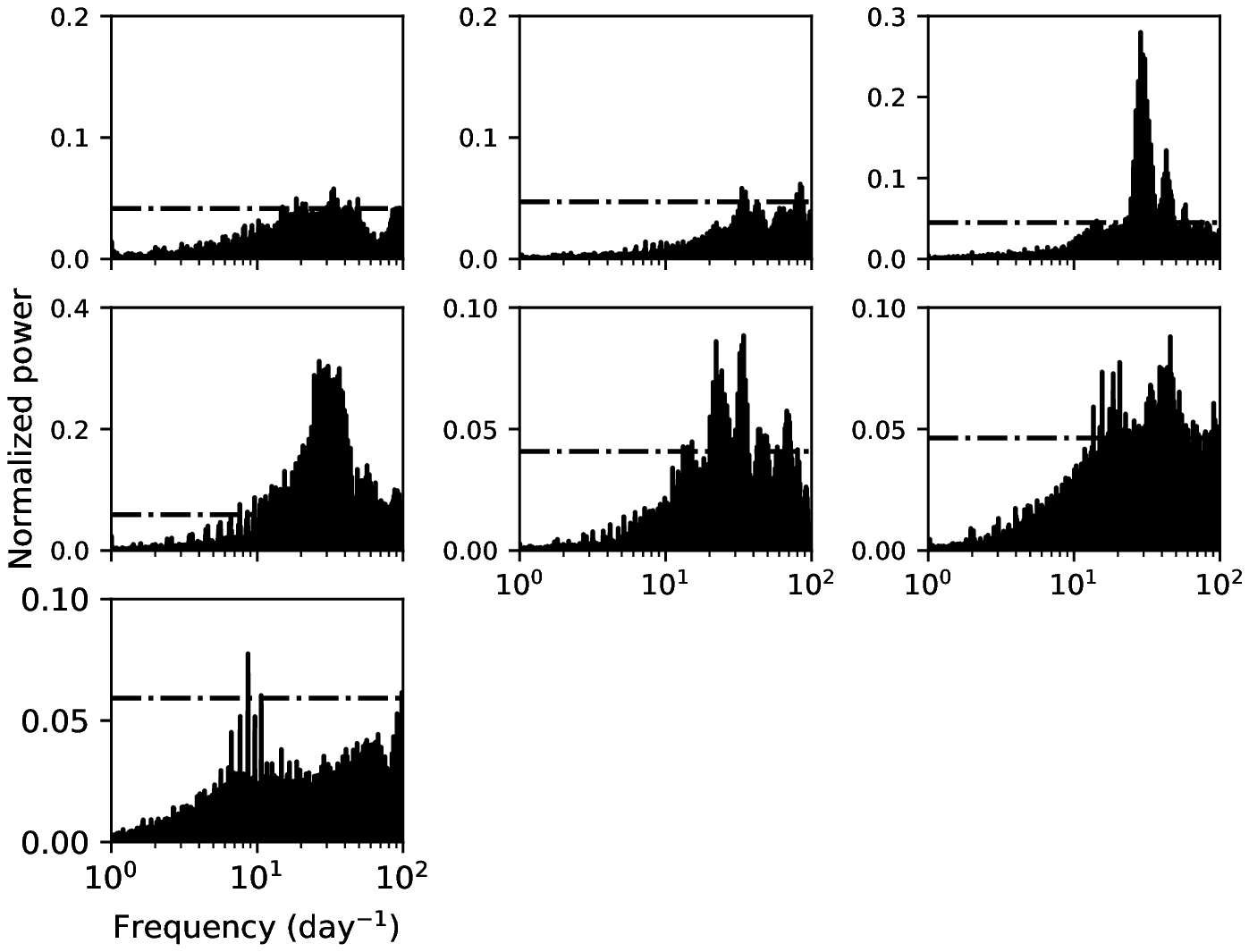}
  }

  \caption{Left : The LS-periodograms of ZTF data for the seven candidates listed in Table~1  with Gaussian and time uncorrelated noise model [the first term in Equation of (\ref{corre})]. The dashed and dotted lines are FAP=0.01 estimated  by the methods of \cite{2008MNRAS.385.1279B}  and of the bootstrap, respectively.
      Right : The LS-periodograms with time correlated noise model [the full expression  of (\ref{corre})].
      The results are for $\tau_{exp}=1$~day, and $\sigma_{exp}=1$~mag, (for G131, G207, G205 and G216)  or  0.5~magnitude (for G453, G432 and G454). The dashed-dotted line represents  FAP=0.01
      estimated from  the method of \cite{2020A&A...635A..83D}.}
   
  \label{ztf}
  \end{figure*}

\section{LS-periodograms and folded light curves of ZTF and TESS data}
\begin{figure*}
\epsscale{1}
\centerline{
  \includegraphics[scale=1]{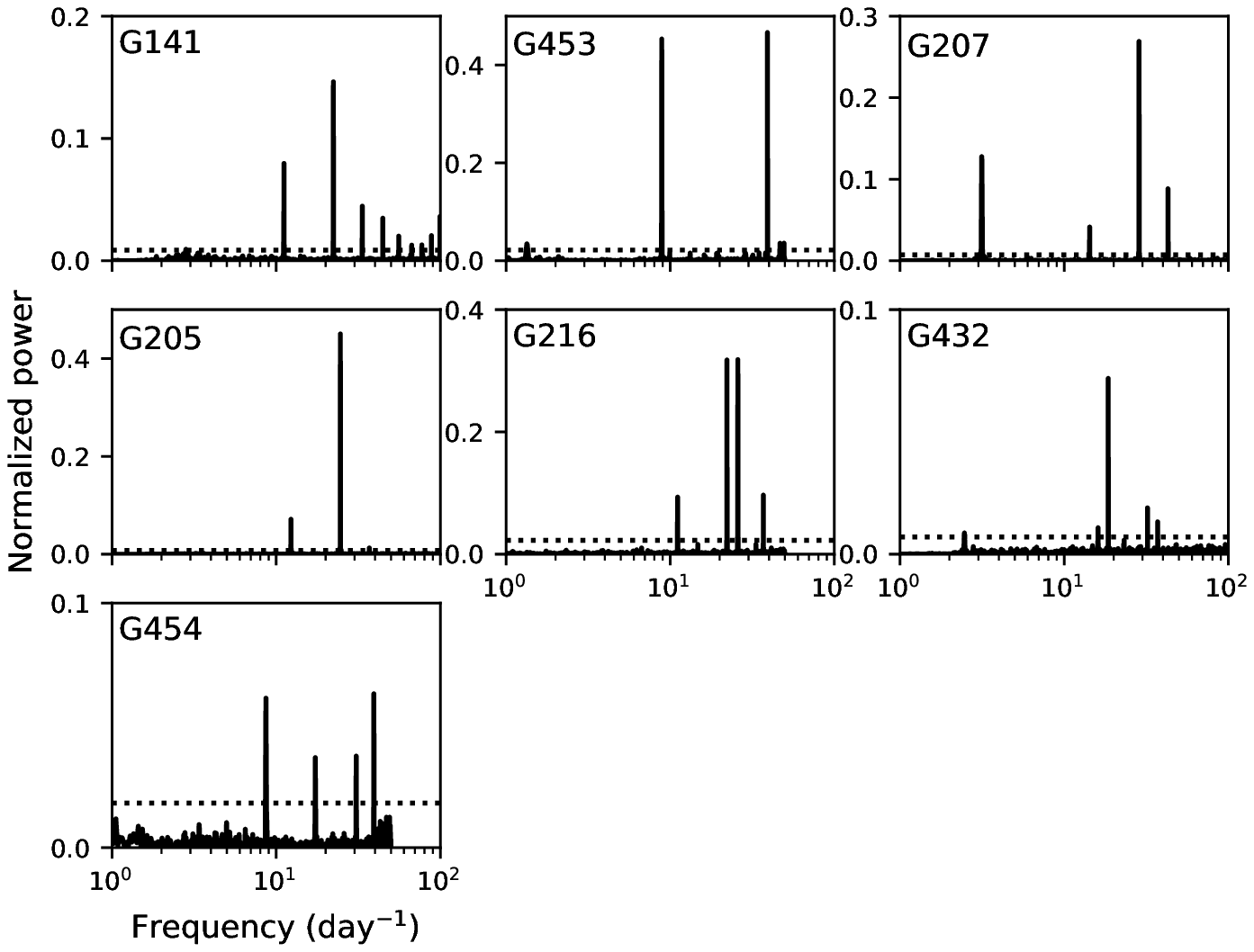}}
\caption{ The LS-periodograms of the TESS data for the seven candidates listed in Table~1 with Gaussian and time uncorrelated noise model [the first term in equation of (\ref{corre})].  The dotted  horizontal lines represent  FAP=0.01 estimated from the bootstrap method.}
\label{tess}
\end{figure*}

\begin{figure*}
  \epsscale{1}
  \centerline{
    \includegraphics[scale=1]{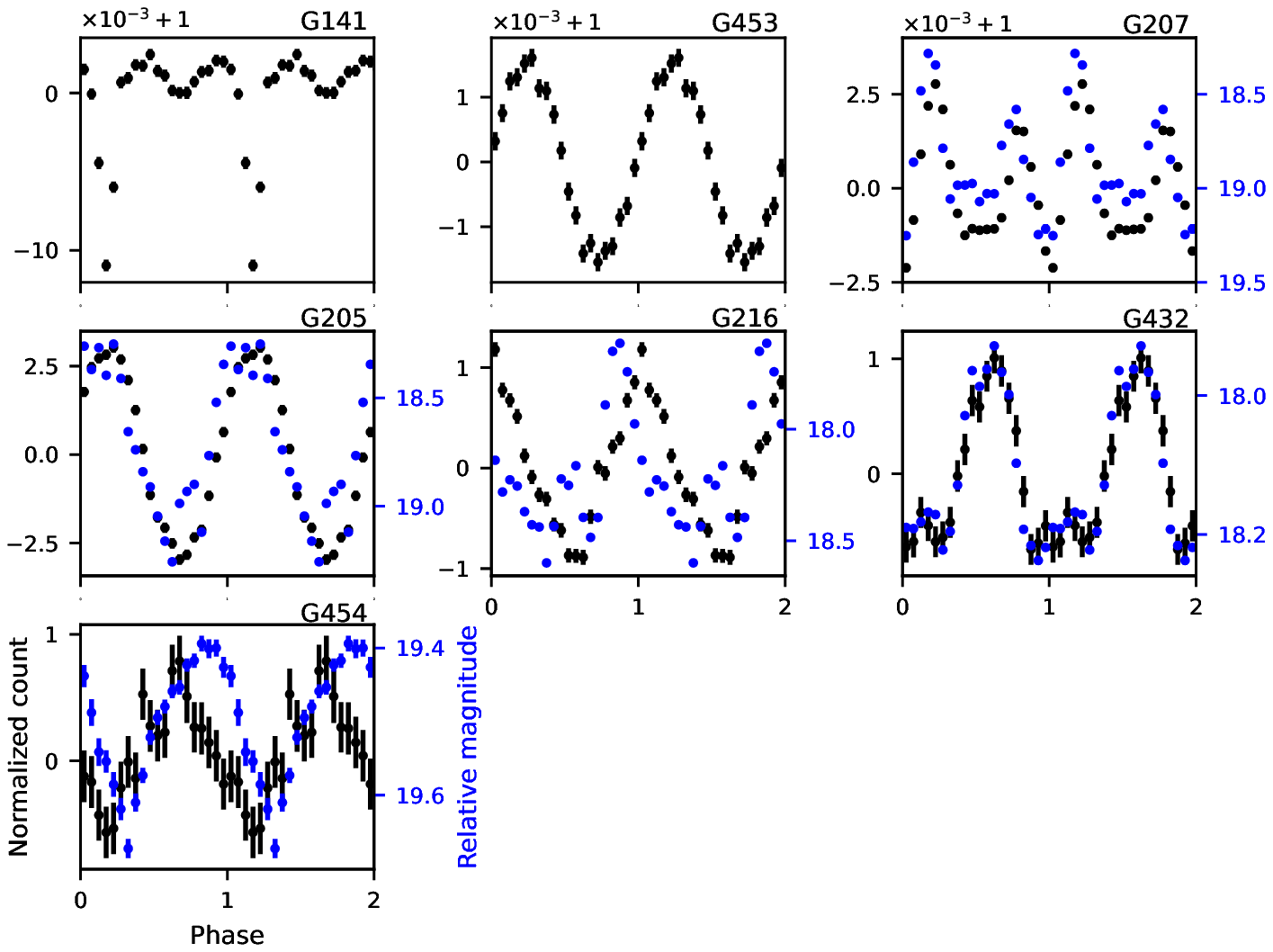}}
  \caption{ZTF (blue color) and TESS (black color) light curves for seven candidates in Table~1. The data is folded with $f_{ZTF}$ (or $F_0$),
    except for $F_0/2$ of G207. For G141 and G453, the ZTF data do not show a periodic signal of $F_0$.  Phase zero corresponds to MJD~59398.713391,
  and two cycles are plotted for clarity.}
  \label{all-light}
\end{figure*}

\begin{figure*}
  \epsscale{1}
  \centerline{
    \includegraphics[scale=0.6]{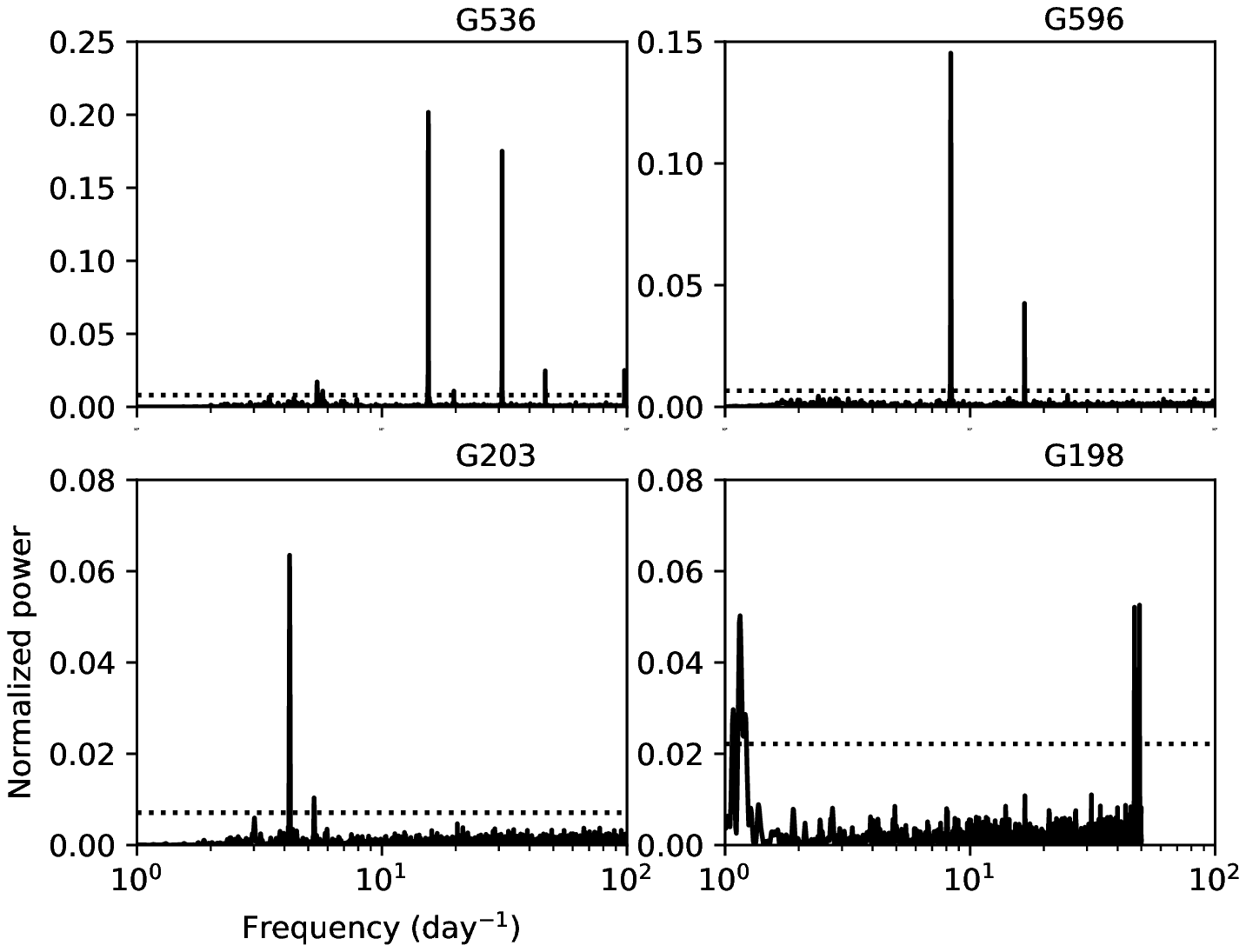}
\includegraphics[scale=0.6]{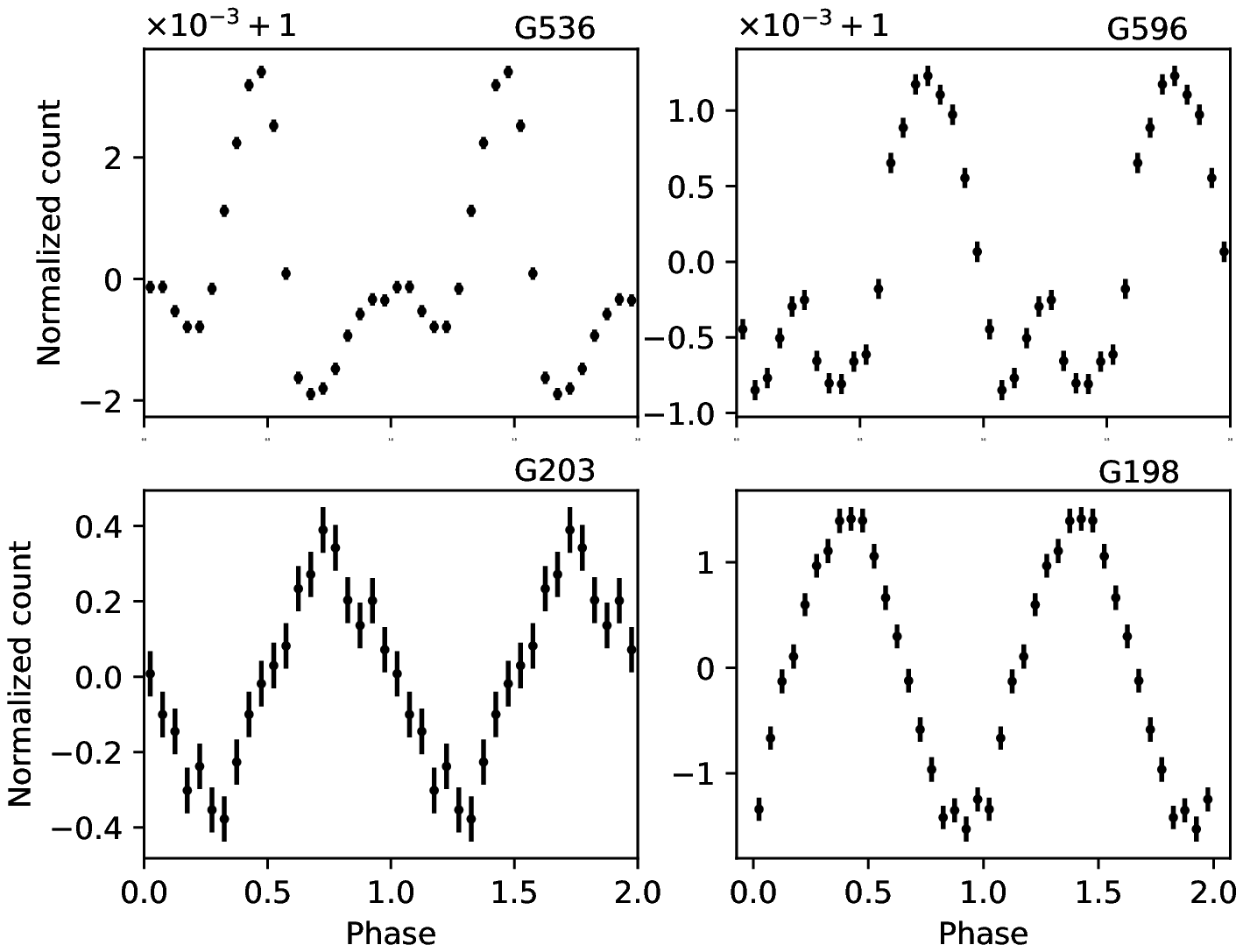}
  }
  \caption{The LS-periodograms (left) and folded light curves (right)  of the TESS data
    for the four candidates listed in Table~4. The dotted lines in the left panel show FAP=0.01 estimated with the bootstrap method. In the right panel, phase zero corresponds to MJD~59398.713391,
    and two cycles are  plotted for clarity.}
  \label{other}
  \end{figure*}

\restartappendixnumbering
Figures~\ref{tess} and \ref{all-light} show  LS-periodograms of the TESS data and folded light curves of ZTF/ TESS data for seven candidates listed in Table~1. Figure~\ref{other} presents the
LS-periodograms and folded light curves of TESS data for the  four candidates listed in Table~4.

\section{Information from the International Variable Star Index and vsnet-chat}
While the article in press, we were noted that the candidates of the orbital periods for  
 4 sources in 7 candidates listed in  in Table~1 have been reported in  
 the international Variable Star Index\footnote{\url{https://www.aavso.org/vsx/index.php?view=search.top}} (VSX, for G205 and G432) and in vsnet-chat (for G141 and G207).  We summarize the information in VSX and vsnet-chat for 7 sources. 

\paragraph{G141,4XMM~J172959.0+522948} This source  is named as  MGAB-V705  in VSX\footnote{\url{https://www.aavso.org/vsx/index.php?view=detail.top&oid=1499054}} 
and the candidate of orbital period $\sim 0.0897008(1)$ day is reported in vsnet-chat~8923\footnote{\url{http://ooruri.kusastro.kyoto-u.ac.jp/mailarchive/ vsnet-chat/8923}}. This value is consistent with the value reported in Table~1. 
\paragraph{G453, 1RXS J185013.9+242222} As noted in main text, this source has been known as a dwarf nova by previous observation of the outburst. This source is named as DDE 163 in VSX\footnote{\url{https://www.aavso.org/vsx/index.php?view=detail.top&oid=686692}}
\paragraph{G207,  2SXPS J195230.9+372016} This source is counterpart of ZTF~18abrxtii, which is listed in VSX\footnote{\url{https://www.aavso.org/vsx/index.php?view=detail.top&oid=2224634}}. The  orbital period $\sim 0.0699896(1)$~day, which is consistent with the value reported in Table~1,  is mentioned in vsnet-chat~8866\footnote{\url{http://ooruri.kusastro.kyoto-u.ac.jp/mailarchive/vsnet-chat/8866}}.
\paragraph{G205, 2SXPS J202600.8+333940} This source is named  as BMAM-V634 in VSX\footnote{\url{https://www.aavso.org/vsx/index.php?view=detail.top&oid=1543125}}, and the candidate of orbital period $\sim$0.0813141~day obtained from  ZTF data \citep{2020ApJS..249...18C} is consistent with our result.
\paragraph{G216, 2SXPS J211129.4+445923} This source is counter part of ZTF~17aaapwae, which is listed in VSX\footnote{\url{https://www.aavso.org/vsx/index.php?view=detail.top&oid=2223038}} as a variable star.
\paragraph{G432, 2SXPS J192530.4+155424} This source is named as DDE 182 in VSX\footnote{\url{https://www.aavso.org/vsx/index.php?view=detail.top&oid=1543028}}. The candidate of the orbital period $\sim 0.053903$~day 
in the catalog is consistent with our result. 
\paragraph{G454, 1RXS J172728.8+132601} As noted in the main text, the outburst happened in 2021 was alerted as a GAIA transient source, AT~2021. ZTF~18abttrrr is listed in VSX\footnote{\url{https://www.aavso.org/vsx/index.php?view=detail.top&oid=2224470}} as a variable star.  

\end{document}